# MAGNETO–SUPERCONDUCTIVITY OF RUTHENO-CUPRATES

V.P.S. AWANA[*]

National Physical Laboratory, K.S. Krishnan Marg, New Delhi-110012, India.

e-mail: awana@mail.nplindia.ernet.in

\* *Work done in part during stay at: Materials and Structures Laboratory, Tokyo Institute of Technology, Yokohama 226-8503, Japan;* ***and*** *Superconducting Materials Center, National Institute for Materials Science, 1-1 Namiki, Tsukuba, Ibaraki 305-0044, Japan.*

## 1. INTRODUCTION

Co-existence of superconductivity and magnetism (in particular the ferromagnetism) has been a point of discussion over decades [1,2]. The two phenomenon being co-existing with each other were realised in various f-electron compounds viz. $ErRh_4B_4$ [3,4]. In more recent years, the phenomenon is observed in $UGe_2$ [5] and $ZrZn_2$ [6]. Various explanations were put forward to understand the phenomenon, viz. accommodation of superconductivity in spiral like magnetic structure [3], or the spin triplet paired superconductivity with spin fluctuating ferromagnetism [3-5]. In fact, according to a long-term common sense superconductivity and magnetic long-range order do not mutually exist within a single (thermodynamical) phase. This is the fundamental reason, which makes the magneto-superconducting compounds altogether more interesting. The topic has been widely discussed in condensed matter physics over decades.

As far as high $T_c$ superconductivity is concerned, the coexistence of high-$T_c$ superconductivity and magnetism was reported for a rutheno-cuprate of the Ru-1222 type, *i.e.* $RuSr_2(Gd_{0.7}Ce_{0.3})_2Cu_2O_{10-\delta}$ [7], and more recently for $RuSr_2GdCu_2O_{8-\delta}$ (Ru-1212) [8]. These reports further renewed the interest in the possible coexistence of superconductivity and magnetism. It is believed that in rutheno-cuprates the $RuO_6$ octahedra in the charge reservoir are mainly responsible both for magnetism and for doping holes into the superconductive $CuO_2$ plane.

The structures of both $RuSr_2GdCu_2O_{8-\delta}$ and $RuSr_2(Gd,Ce)_2Cu_2O_{10-\delta}$ are derived from that of $RBa_2Cu_3O_{7-\delta}$ or $CuBa_2RCu_2O_{7-\delta}$ with Cu in the charge reservoir replaced by Ru such that the $CuO_{1-\delta}$ chain is replaced by a $RuO_{2-\delta}$ sheet [9]. In the $RuSr_2(Gd,Ce)_2Cu_2O_{10-\delta}$ structure furthermore, a three-layer fluorite-type block instead of a single oxygen-free $R$ (= rare earth element) layer is inserted between the two $CuO_2$ planes of the Cu-1212 structure. Schematic unit cell of both Ru-1222 and Ru-1212 are shown below in Figure 1.

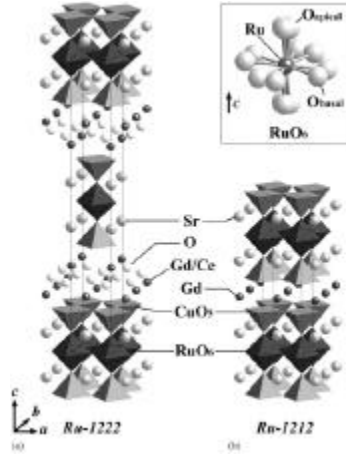

**Fig.1**. Schematic diagrams of (a) Ru-1212 and (b) Ru-1222. The inset is schematic diagram for the average structure of RuO$_6$ octahedra.

Although bulk magnetism due to ordering of the Ru moments was confirmed for Ru-1212 from μSR (muon spin rotation/relaxation/resonance) and ESR (electron spin resonance) studies [8,10], the exact type of ordering is still debated [11]. In particular, the results revealed from neutron scattering experiments [11,12] and magnetization studies [8, 13-15] do not agree with each other. While the former concludes the ordering to be of antiferromagnetic nature, the latter indicates some ferromagnetic ordering. In fact even the most recent magnetization results have casted some doubts on earlier studies related to magnetism of Ru-1212 compounds [16,17]. Worth mentioning is the fact that the so-called ferromagnetic transition at around 140 K is not seen thermodynamically in heat capacity ($C_P$) measurements, only a broad hump extended until room temperature over a length of around 150 K is seen [15].

One of the puzzles in understanding the magnetization data of Ru-1212, is not only the antagonising nature of superconducting and magnetic order parameters, but also the presence of magnetic Gd (8μB) along with the possibly magnetic Cu moment (Cu is paramagnetic in under-doped HTSC), which hinders in knowing the exact Ru spins magnetic contribution to the system. Ru-1212 can be formed with non-magnetic RE (rare earth) Y instead of Gd, but only with HPHT (high pressure high temperature) synthesis technique [12,18]

The bulk nature of superconductivity in these compounds was initially criticised due to lack of Meissner fraction in Gd/Ru-1212 and rather a crypto superconducting phase was proposed [19]. Appearance of bulk superconductivity in Ru-1212 is also confirmed from specific heat ($C_P$) measurements [20,21], though the existing reports do not agree with each other in terms of $C_P$ measurements under magnetic field [$C_p(H, T)$]; in particular the $T_c$ as viewed from the $C_P$ peak increases with field in one report [20] but decreases with field in another [21]. While the former indicates towards the triplet pairing, the latter suggests a normal under-doped HTSC case. It is also suggested that small impurity of GdSr$_2$RuO$_6$ presumably present in the samples of Ref. 20 is responsible for the $C_P(H, T)$ behaviour being different from that of Ref. 21 [22]. It seems that existing reports on the magneto-superconductivity in Gd/Ru-1212 do not agree completely with each other.

As far as Ru-1222 is concerned, the main features are the same as for both Ru-1212. The magnetic structure of Ru-1222 has been studied by neutron powder diffraction [23]. Despite the fact that various physical-property measurements have been carried out on Ru-1212 [8-22] and Ru-1222 [7,23-28], no final consensus has been reached, *i.e.* discussion on their basic characteristics in terms of the oxygen stoichiometry, valence state of Ru, carrier concentration and doping mechanism has not been completed yet. This becomes more important in the event when contradictory experimental results are obtained on different samples, as discussed above [20-22,29,30]. Also, it has been reported [30] that solid solutions of composition $(Ru_{1-x}Cu_x)Sr_2GdCu_2O_{8-\delta}$ can form within $0 < x < 0.75$ with $T_c$ up to 74 K. Interestingly with the higher $x$ values in the above composition the Ru spins do not order magnetically down to 5 K. Henceforth to conclude the coexistence of long-range magnetic ordering of Ru spins with superconductivity in the CuO$_2$ plane, one should strictly avoid the formation of $(Ru_{1-x}Cu_x)$-1212 solid solutions in pristine Ru-1212. Worth mentioning is the fact that Ru/Cu intermixing becomes more complicated as the two elements cannot be distinguished without ambiguity by neutron diffraction, a technique commonly used for fixing various cation occupancies in inorganic solids. Both Ru and Cu do have nearly the same scattering cross-sections for thermal neutrons. The concern of phase purity at the microscopic level in both Ru-1212 and Ru-1222 still remains unresolved. Also, we should look more carefully at the existing contradictions in the reported literature on rutheno-cuprates. Nevertheless, results of recent NMR (nuclear magnetic resonance) experiments on Ru-1212 were interpreted in terms of the coexistence of superconductivity and magnetism [31]. In current review, not only the existing literature is critically accessed, but also very recent data in terms of phase formation and structural, thermal, magnetic, electrical, spectroscopic and microscopic characterization for both Ru-1212 and Ru-1222 will be presented. Recent advancements regarding HPHT high pressure high temperature phase formation of both Ln/Ru-1212 and Ln/Ru-1222 (Ln = Lanthenides) is given along with their physical characterization. Also phase formation of higher derivatives of Ru-1212 and Ru-1222 families viz. Ru-1232 will be discussed. Instead of sticking to vastly studied Ln = Gd systems, the recent results for other lanthanide ruthenocuprates will be discussed. It is further stated that co-existence of superconductivity and ferromagnetism in these compounds is yet far from conclusive.

## 2. EXPERIMENTAL DETAILS

### (2.a) Normal Pressure High Temperature (NPHT) Synthesis

Samples of $RuSr_2GdCu_2O_{8-\delta}$ and $RuSr_2(Gd_{0.75}Ce_{0.25})_2Cu_2O_{10-\delta}$ were synthesized through a solid-state reaction route from stoichiometric amounts of $RuO_2$, $SrO_2$, $Gd_2O_3$, $CeO_2$ and CuO. Calcinations were carried out on mixed powders at 1000 °C, 1020 °C and 1040 °C for 24 hours at each temperature with intermediate grindings. The pressed bar-shaped pellets were annealed in a flow of oxygen at 1075 °C for 40 hours and subsequently cooled slowly over a span of another 20 hours down to room temperature. These samples are termed as "as-synthesized". Part of the as-synthesized samples were further annealed in high-pressure oxygen (100 atm) at 420 °C for 100 hours and subsequently cooled slowly to room temperature. These samples are termed as "100-atm $O_2$-annealed". Further some of the samples were treated in flow of $N_2$ gas at 420 °C for 24 hours and subsequently cool down to room temperature in same gas atmosphere. Though the heat treatments used for the samples in our study are in general similar to those as reported in literature [7-11, 13-15, 17, 19-30], minor differences do exist from one laboratory to another in terms of annealing hours and the temperatures used. Also, it has been reported that not always all samples of the same batch with similar heating schedule show superconductivity [19,22,29]. Our general experience is also the same particularly for Ru-1212, in which achieving superconductivity seems to be a tricky job.

### (2.b) High Pressure High Temperature (HPHT) Synthesis

Worth mentioning is the fact that for both Ru-1212 and Ru-1222, single-phase samples are achieved only for $R$ = Gd, Sm and Eu, with the normal heating schedules mentioned above. For $R$ = Y and Dy, *etc.*, one needs to employ the HPHT (high-pressure high-temperature) procedure for attaining the Ru-1212 phase [12,16,18,31]. In case of Ln/Ru-1212 compounds starting materials for high-pressure synthesis were $RuO_2$ (99.9%), $SrO_2$, $SrCuO_2$, $Ln_2O_3$ (99.9) and CuO (99.9%). These materials were mixed in an agate mortar to obtain starting mixtures for high-pressure synthesis. About 300 mg of starting mixture was sealed in a gold capsule and allowed to react in a flat-belt-type high-pressure apparatus at 6 GPa, and at 1200–1300°C for 3 h, then quenched to room temperature. For Ln = Y system, high-purity sample was obtained with the `Ru-poor' starting composition, of $Ru_{0.9}Sr_2YCu_2O_{7.8}$. A similar procedure is followed in ref. [12,16,18,31]

For Ln/Ru-1222, the samples of composition $RuSr_2(Ln_{3/4}Ce_{1/4})_2Cu_2O_{10}$ with Ln = Ho, Y and Dy were synthesised through a HPHT solid-state reaction route. For the HPHT synthesis and to fix the oxygen at 10.0 level, the molar ratio used were: $(RuO_2) + (SrO_2) + (SrCuO_2) + 3/4(CuO) + 1/4(CuO_{0.011}) + 3/4(Ln_2O_3) + 1/2(CeO_2)$ resulting in $RuSr_2(Ln_{3/4}Ce_{1/4})_2Cu_2O_{10}$. $CuO_{0.011}$ is pure Cu-metal, for which precise oxygen content is determined before use. The materials were mixed in an agate mortar. Later around 300 mg of the mixture was sealed in a gold capsule and allowed to react in a flat-belt-type-high-pressure apparatus at 6GPa and 1200 °C for 2 hours [32]. Nearly no change was observed in the weight of synthesized samples, indicating towards their fixed nominal oxygen content.

We believe the oxygen content of all the samples is close to nominal i.e. 10. Determination of the oxygen content of synthesized samples is yet warranted to know the oxygen value for these samples. In case of Ru-1232 the samples of composition $RuSr_2(Ln_1Ce_2)Cu_2O_{12.25}$ with Ln = Y and Dy were synthesised through a HPHT solid-state reaction route. For the HPHT synthesis, the molar ratio used were: $(RuO_2) + (SrO_2) + (SrCuO_2) + 3/4(CuO) + 1/4(CuO_{0.011}) + 1/2(Ln_2O_3) + 2(CeO_2)$ resulting in $RuSr_2(Ln_1Ce_2)Cu_2O_{12.25}$. $CuO_{0.011}$ is pure Cu-metal, for which precise oxygen content is determined before use. The materials were mixed in an agate mortar. Later around 300 mg of the mixture was sealed in a gold capsule and allowed to react in a flat-belt-type-high-pressure apparatus at 6GPa and 1200 °C for 2 hours. Nearly no change was observed in the weight of synthesized samples, indicating towards their fixed nominal oxygen content. We believe the oxygen content of both the samples is close to nominal i.e. 12.25.

### (2.c) Physical property experimentation for NPHT and HPHT samples.

Thermogravimetric (TG) analyses (Perkin Elmer: System 7) were carried out in a 5 % $H_2$/95 % Ar atmosphere at the rate of 1 °C/min to investigate the oxygen non-stoichiometry. X-ray diffraction (XRD) patterns were collected at room temperature (MAC Science: MXP18VAHF[22]; Cu$K_\alpha$ radiation). Magnetization measurements were carried out on a superconducting-quantum-interference-device (SQUID) magnetometer (Quantum Design: MPMS-5S). Resistivity measurements under an applied magnetic field of 0 - 7 T were performed in the temperature range of 5 - 300 K using a physical-property-measurement system (Quantum Design: PPMS). Electron diffraction (ED) patterns were taken at room temperature using an analytical transmission electron microscope (Hitachi: HF-3000S) with a cold field emission gun operated at an accelerating voltage of 300 kV. The SAED and CBED patterns were taken from specimen areas of about 100 and 8 nm, respectively. The HREM images were taken using a high-resolution, high-voltage transmission electron microscope (Hitachi: H-1500) operated at an accelerating voltage of 800 kV. The Ru $L_{III}$-edge XANES measurements were performed at room temperature for polycrystalline samples at the BL15B beamline of the Synchrotron Radiation Research Center (SRRC) in Hsinchu, Taiwan.

## 3. RESULTS & DISCUSSION

### 3.1 Phase formation and lattice parameters: X-ray diffraction results for NPHT Gd/Ru-1212 and Gd/Ru-1222 samples

$RuSr_2GdCu_2O_{8-\delta}$ (Ru-1212) samples possess a tetragonal Ru-1212 structure with a space group $P4/mmm$. Lattice parameters were determined at $a = b = 3.8218(6)$ Å and $c = 11.476(1)$ Å. Corresponding X-ray diffraction pattern is shown in Fig. 2. Small amount of $SrRuO_3$ is also seen, which is marked on the pattern [33].

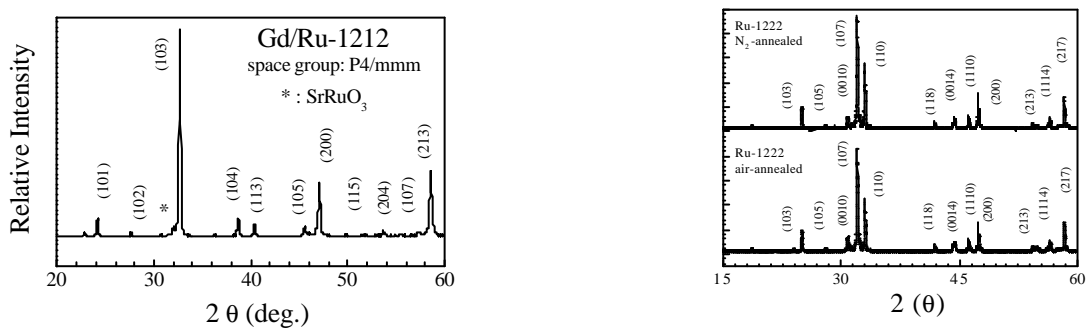

**Figs. 2&3.** X-ray diffraction pattern for an as-synthesized $RuSr_2GdCu_2O_{8-\delta}$ sample & for an as-synthesized and $N_2$ annealed $RuSr_2(Gd_{0.75}Ce_{0.25})_2Cu_2O_{10-\delta}$ sample.

Essentially no difference is found in X-ray diffraction pattern or lattice parameters for the 100-atm $O_2$-annealed Ru-1212 sample, suggesting that the oxygen content remained unchanged upon the high-$O_2$-pressure annealing. Worth mentioning is the fact that earlier phase-pure (no trace of $SrRuO_3$ within the XRD detection limit) Ru-1212 sample was not superconducting, even with various types of post-annealing treatments [29]. Compared with Ru-1212, the Ru-1222 phase forms more easily in impurity-free form. Both the as-synthesized and the 100-atm $O_2$-annealed Ru-1222 samples were found to be of high quality in terms of phase purity. An X-ray diffraction pattern for an as-synthesized sample is shown in Fig. 3. The lattice parameters were determined from the diffraction data in the tetragonal space group $I4/mmm$: $a = b = 3.8337(6)$ Å and $c = 28.493(1)$ Å for the as-synthesized sample, and $a = b = 3.8327(7)$ Å and $c = 28.393(1)$ for the 100-atm $O_2$-annealed sample [34]. The shorter lattice parameters for the 100-atm $O_2$-annealed sample are believed to manifest the fact that it is more completely oxygenated than the as-synthesized sample. For $N_2$ – annealed sample the lattice parameters are $a = b = 3.8498(3)$ Å and $c = 28.4926(9)$ Å [35]. An increase in lattice parameters of $N_2$-annealed sample indicates an overall decrease in oxygen content of the sample. X-ray diffraction results, in terms of phase purity, space groups and the obtained lattice parameters are in general accordance with the reported literature for both Gd/Ru-1212 and Gd/Ru-1222 variously annealed samples.

### 3.1 Phase formation and lattice parameters: X-ray diffraction results for HPHT Ln/Ru-1212, Ln/Ru-1222 and Ln/Ru-1232 samples

Phase formation of Ln/Ru-1212 compounds is reported in ref.[18]. Nearly single-phase materials were obtained for various Ln, but with slight off stoichiometry of Ru. For example in case Y/Ru-1212 the single phase formation is achieved with nominal composition of $Ru_{0.9}Sr_2YCu_2O_{7.8}$ [18]. In case of Ln/Ru-1222 samples with composition $RuSr_2(Ln_{3/4}Ce_{1/4})_2Cu_2O_{10}$ with Ln = Ho, Y and Dy were crystallised in a single-phase form in space group $I4/mmm$ with lattice parameters $a = b = 3.819(1)$ Å, and $c = 28.439(1)$ Å for Ln = Y, $a = b = 3.813(2)$ Å, and $c = 28.419(1)$ Å for Ln = Ho, and $a = b = 3.824(4)$ Å, and $c = 28.445(1)$ Å for Ln = Dy. The volume of the cells is 413.2, 414.8 and 415.9 Å$^3$ for Ln = Ho, Y and Dy respectively. The trend of their cell volumes is in line with the rare earths ionic sizes. Figure 4 shows the X-ray diffraction patterns of finally synthesized Ln/Ru-1222 compounds. As seen from this figure these compounds are crystallised in a single-phase form with only small amount of $SrRuO_3$ present in Ln = Y sample [32].

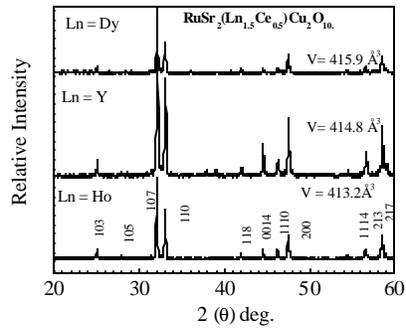

**Fig. 4.** X-ray diffraction pattern for Ln/Ru-1222.

In Case of Ln/Ru-1232, the X-ray diffraction patterns of $RuSr_2(Y_1Ce_2)Cu_2O_{12+\delta}$ with $\delta = 0.0$, 0.10 and 0.20 which could be readily indexed within tetragonal structure having space group $P4/mmm$. Also seen are small quantities of $SrRuO_3$ and $RuSr_2(RE_{1.5}Ce_{0.5})Cu_2O_{10}$ (Ru-1222). Though, all the three oxygen contents gave nearly similar X-ray patterns, we decided to work with $\delta = 0.25$, with a pre-assumption that higher oxygen content could give rise to better superconductivity. Lattice parameters calculated are $a = b = 3.822(1)$ Å, and $c = 16.3336(4)$ Å for Ln = Y and $a = b = 3.827(3)$ Å, and $c = 16.3406(7)$ Å for Ln = Dy samples of the $RuSr_2(Ln_1Ce_2)Cu_2O_{12.25}$ series. The XRD patterns for $RuSr_2(Y_1Ce_2)Cu_2O_{12+\delta}$ with $\delta = 0.0$, 0.10 and 0.20 are shown in Fig. 5.

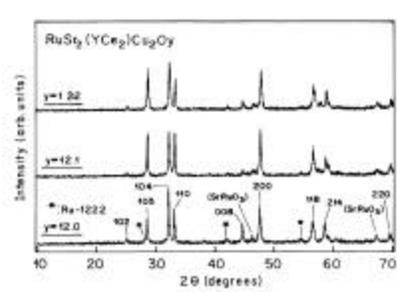

**Fig. 5.** X-ray diffraction pattern Ln/Ru-1232 with $\delta = 0.0$, 0.10 and 0.20

### 3.2 Oxygen stoichiometry: TG results

Thermogravimetric reduction curves were recorded for both as-synthesized and 100-atm $O_2$-annealed samples of Gd/Ru-1212 and Gd/Ru-1222 [36]. Two examples of the obtained TG curves are shown in Fig. 6. For both the phases the decomposition of sample occurs in two distinct steps about 200–350 and 400–500 °C.

To clarify the process of the sample decomposition during the different steps of reduction, additional TG measurements were performed in the same atmosphere for the simple oxides of the constituent metals, i.e. $RuO_2$, CuO, $Cu_2O$ and $CeO_2$. It was seen that $RuO_2$, CuO and $Cu_2O$ decompose into corresponding metals at low temperatures of about 100, 150 and 330 °C, respectively, while $CeO_2$ remains stable at least up to 750 °C [36]. We therefore calculated the exact oxygen contents of the Ru-1212 and Ru-1222 samples from the weight losses seen in the H2/Ar reduction curves by 550 °C assuming the final decomposition product to be a mixture of oxides, SrO, $Gd_2O_3$ and $CeO_2$, and Ru and Cu metals.

The results are presented in Table 1, implying that the 100-atm $O_2$-annealed Ru-1212 sample is stoichiometric within the error bars of the analysis, while the 100-atm O2-annealed Ru-1222 sample is clearly oxygen deficient. Furthermore, as already assumed based on the lattice parameters and the Tc values, the difference between the 100-atm and 1-atm O2-annealed samples is larger for the Ru-1222 phase than for the Ru-1212 phase, see also the data for relevant samples in Refs. [25,33,34]. In Table 1, we also show the absolute oxygen content values for the samples annealed in Ar up to the highest temperatures before the break-down of the structure, i.e. 750 °C for Ru-1212 and 900 °C for Ru-1222. These values, i.e. 7.80(5) for Ru-1212 and 9.35(5) for Ru-1222, represent the minimum oxygen contents tolerated by these phases.

Table 1: The value of oxygen content as determined from TG analysis for variously treated Gd/Ru-1212 and Gd/Ru-1222 samples.

| Synthesis/annealing conditions | RuSr$_2$GdCu$_2$O$_{8-d}$ (Ru-1212) | RuSr$_2$Gd$_{1.5}$Ce$_{0.5}$Cu$_2$O$_{10-d}$ (Ru-1222) |
|---|---|---|
| 100-atm O$_2$, 420 $^0$C | 7.98(5) ($T_c$ = 20 K) | 9.63(5) ($T_c$ = 43 K) |
| 1-atm O$_2$, 1075 $^0$C | 7.93(5) ($T_c$ = 20 K) | 9.54 (5) ($T_c$ = 23 K) |
| 1-atm Ar, 750/900 $^0$C (Ru-1212/Ru-1222) | 7.80 (5) (No $T_c$) | 9.35(5) (No $T_c$) |

As far as the HTHT treated various Ln/Ru-1212, 1222 σ 1232 compounds are concerned their oxygen stoichiometry is believed to be very close to nominal, because nearly no change was observed in the weight of synthesized samples, indicating towards their fixed nominal oxygen content. We believe the oxygen content of all the samples is close to nominal. Determination of the oxygen content of the HPHT synthesized samples is yet warranted.

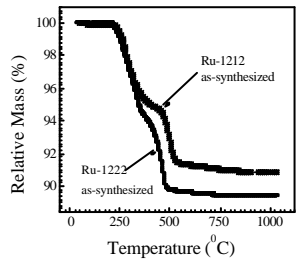

**Fig. 6.** TG curves for as-synthesized Gd/Ru-1212 and 1222 samples recorded in 5 % H$_2$/95 % Ar atmosphere.

**3.3 Valence of Ru: XANES spectroscopy results**

The Ru $L_{III}$-edge XANES spectra were obtained for an as-synthesized and a 100-atm O$_2$-annealed Ru-1222 samples [37]. The spectra were analyzed quantitatively by fitting them to certain linear combinations of those for reference materials Sr$_2$RuO$_4$ (Ru$^{IV}$) and Sr$_2$GdRuO$_6$ (Ru$^V$). All the spectra exhibited two peaks, the higher-energy one and the lower-energy one being due to $2p \rightarrow e_g$ and $2p \rightarrow t_{2g}$ transitions, respectively [38,39]: with increasing Ru valence from +IV to +V, the crystal-field splitting increases and thereby the separation between the two peaks enhances. Furthermore, the peaks were accordingly shifted by ~1.5 eV to the higher energy. It was concluded that both Ru-1222 samples are between the two reference materials in terms of the Ru valence. Fitting of the spectra revealed a valence value of +4.74 for the as-synthesized sample and +4.81 for the 100-atm O$_2$-annealed sample [37]. The obtained result suggests that the valence of Ru in Ru-1222 is affected by the change in oxygen content. It is therefore interesting to compare the presently obtained Ru valence values to that previously reported for a RuSr$_2$(Gd$_{0.7}$Ce$_{0.3}$)$_2$Cu$_2$O$_{10-\delta}$ sample (+4.95) with $T_c \approx 60$ K [38]. For the three Ru-1222 samples the Ru valence/$T_c$ values thus were: +4.74/30 K, +4.81/43 K and +4.95/60 K. (Note that the XANES measurements and analyses were carried out in parallel ways for all the three samples.) The latter two samples were both annealed under 100 atm oxygen pressure, but with different temperature programs. It is thus likely that the one previously reported [38] had somewhat higher oxygen content than the present 100-atm O$_2$-annealed sample. It seems that the valence of Ru in Ru-1222 depends on the oxygen content, thus indirectly suggesting that the changes in oxygen stoichiometry occur in the RuO$_{2-\delta}$ layer. Here it is interesting to note that a previous study had shown that Ru remains essentially unchanged (close to pentavalent) upon varying the Ce-substitution level within $0.3 \leq x \leq 0.5$ in fully oxygen-loaded RuSr$_2$(Gd$_{1-x}$Ce$_x$)$_2$Cu$_2$O$_{10-\delta}$ samples [38,39].

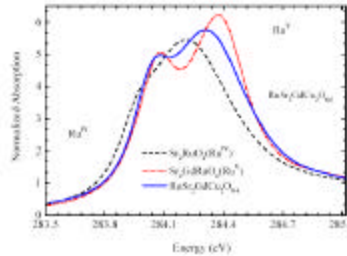

**Fig. 7.** Ru $L_{III}$-edge XANES spectra for reference materials, Sr$_2$RuO$_4$ (Ru$^{IV}$) and Sr$_2$GdRuO$_6$ (Ru$^V$), and for as-synthesized RuSr$_2$GdCu$_2$O$_{8-\delta}$ sample.

Ru $L_{III}$-edge XANES analysis for an as-synthesized (final annealing in $O_2$ at 1075 $^oC$) $RuSr_2GdCu_2O_{8-\delta}$ sample revealed a valence value of +4.62 [36], see Fig.7. Same analysis procedure as discussed above was followed. This value is very close to that reported in Ref. [39] for a $RuSr_2GdCu_2O_{8-\delta}$ sample, *i.e.* +4.60, with the final annealing performed in $O_2$ at 1060 $^oC$.

### 3.4 Superstructures: SAED/HRTEM results

**3.4.a SAED for NPHT synthesized Gd/Ru-1222 and Gd/Ru/1212 samples**

Figs. 8(a)–(d) show SAED patterns of the Ru-1222 sample taken with [0 0 1], [1 0 0], [1 1 0] and [3 1 0] incidence, respectively. Main reflections of h k l can be indexed by the fundamental lattice, and the sharp reflections are seen in Figs. 2(b) and (c). On the other hand, superlattice reflections with diffuse streaks along the c* direction are seen in Fig. 2(d) as indicated by white arrowheads. It has been confirmed that the superlattice reflections are corresponding to those indicated by the white arrowheads in the [0 0 1] SAED pattern (Fig. 2(a)). The diffuse streaks along the c* direction suggest existence of domain structures or stacking disorders of the superlattice in the c direction [40].

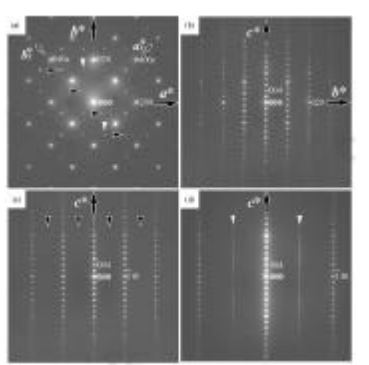

**Fig. 8**. SAED patterns of Ru-1222. h k l and h k lS are corresponding to the indexes of Reciprocal fundamental lattice (a*, b*, c*) and superlattice ($a_S^*, b_S^*, c_S^*$), respectively: (a) [0 0 1] SAED pattern; (b) [1 0 0] SAED pattern; (c) [1 1 0] SAED pattern; and (d) [3 1 0] SAED pattern.

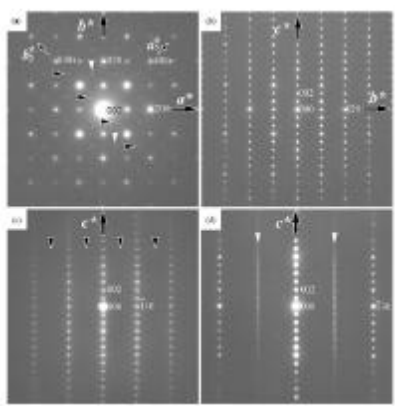

**Fig. 9.** SAED patterns of Ru-1212: (a) [0 0 1] SAED pattern; (b) [1 0 0] SAED pattern; (c) [1 1 0] SAED pattern; and (d) [3 1 0] SAED pattern.

Similar sets of SAED patterns of Ru-1212 are shown in Figs. 9(a)–(d) for [0 0 1], [1 0 0], [1 1 0] and [3 1 0] incidence, respectively, although main reflections of h k l can be indexed by the fundamental primitive tetragonal lattice which has already been reported as the 1212-type structure (P4/mmm: $a$ = 0.38337(6) nm, $c$ = 11.4926(9) nm) [259]. The superlattice reflections, which are corresponding to those caused by a supercell, are also observed in Fig. 3(a) as indicated by the white and black arrowheads [41]. It should be noted that the superlattice reflections with diffuse streaks along c* direction are also observed in Fig. 3(d) as indicated by white arrowheads. Therefore, existence of domain structures or stacking disorders of the superlattice as seen in Ru-1222 are also expected in Ru-1212.

**3.4.b HRTEM for NPHT synthesized Gd/Ru-1222 and Gd/Ru/1212 samples**

Fig. 10 shows a HREM (high resolution transmission electron microscopy) image of Gd/Ru-1222 taken with [1 0 0] incidence. The layers indicated by Ru, Sr and Cu are assumed to be corresponding to those of $RuO_2$, SrO and $CuO_2$, respectively. The two layers indicated by Gd and/or Ce are corresponding to parts of the fluorite-type block of $(Gd_{1.5}Ce_{0.5})_2O_2$. It is confirmed that the layers and the fluorite-type block are stacked along the c-axis following 1222-type structure without any intergrowths [42]. Fig. 11 shows a HREM image of Gd/Ru-1212 taken with [1 0 0] incidence. The layers indicated by Ru, Sr, Cu and Gd are assumed to be corresponding to those of $RuO_2$, SrO, $CuO_2$ and Gd, respectively. It is confirmed that the layers are stacked along the c-axis following 1212-type structure without any intergrowths [42].

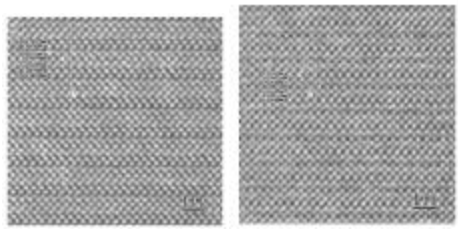

**Fig. 10 & 11.** HREM images of Gd/Ru-1222 and Gd/Ru-1212 being taken at [1 0 0] incidence. No stacking disorders due to intergrowths are observed.

**3.4.c. Structural model based on SAED and HRTEM results for Gd/Ru-1222**

Fig. 12(a) shows a schematic diagram of the fundamental structure of Ru-1222 in which only $RuO_2$ planes are shown as indicated by filled squares. The rectangular parallelepiped surrounded by solid lines represents the fundamental body-centered tetragonal lattice. The rectangular parallelepipeds surrounded by the dotted lines are corresponding to the superlattice. It is assumed on the basis of the result by Knee et al. [23] that the superlattice reflections observed in this electron diffraction study are due to an ordering of the $RuO_6$ octahedra rotated about the c-axis. Fig. 12(b) shows a schematic diagram of the ordering of rotations about the c-axis, that form the super-cell. The rectangular parallelepipeds surrounded by the solid lines and dotted lines are corresponding to the fundamental lattice and superlattice, respectively. Dark- and light-gray squares indicate right- and left-handed rotations of the RuO6 octahedra about the c-axis, respectively.

It is confirmed that there are two possible arrangements, A and B, of the rotated $RuO_6$ octahedra as indicated by arrowheads in Fig. 12(b) if we consider the rotative directions of the $RuO_6$ octahedra located at the body-center of the fundamental lattice of Fig. 12(a). The rotated $RuO_6$ octahedra in both cases of A and B are ordered along the c direction with an interval of c that is the same as that of the fundamental lattice of Fig. 12(a). It is confirmed that the lattice types of A and B can be considered as A- and B-centered orthorhombic superlattices (A and B superlattices), respectively, if the axes of $a_S$ and $b_S$ are fixed as shown in Fig. 12(b).

It should be noted that the A and B superlattices are crystallographically identical to each other, being mutually related by 90° rotation about the c-axis. However, we focus here on the A and B superlattices in order to interpret the experimental diffraction patterns. Fig. 12(c) shows a schematic diagram of an example of our proposed domain structure model for explaining the experimental results in this study. That is, Ru-1222 is composed of superlattice domains of about 10 nm in width distinguished by the A and B superlattices along the c direction as shown in Fig. 12(c).

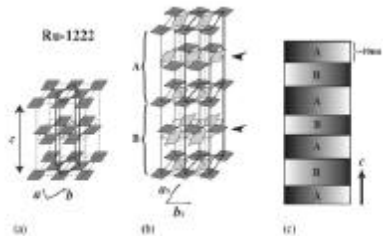

**Fig. 12.** Schematic diagrams of the fundamental lattices and superlattices of Ru-1222 the fundamental lattice of Ru-1222, only $RuO_2$ planes are shown (filled squares); (b) the two possible superlattices (A and B superlattices) of Ru-1222; (c) an example of a proposed domain-structure model for Ru-1222.

### 3.4.d. Structural model based on SAED and HRTEM results for Gd/Ru-1212

Fig. 13(a) shows a schematic diagram of the fundamental primitive tetragonal lattice of Ru-1212. It is assumed on the basis of the results by McLaughlin et al. [41] that the superlattices, which form the super-cell, are due to ordering of the $RuO_6$ octahedra rotated about the c-axis. We consider here two possible arrangements, P and I, of the rotated $RuO_6$ octahedra as indicated by arrowheads in Fig. 13(b) by analogy with the superlattice models of Ru-1222. The rotated $RuO_6$ octahedra of P and I are ordered along the c direction with intervals of c and 2c, respectively. The different periodicities between P and I are caused by the differences of rotative direction of the $RuO_6$ octahedra indicated by arrowheads in Fig. 13(c). It is confirmed that the lattice types of P and I are primitive and body-centered tetragonal superlattices (P and I superlattices), respectively. It should be noted that the P and I superlattices of Ru-1212 are crystallographically different to each other, while the A and B superlattices of Ru-1222 are crystallographically identical to each other. We propose a domain structure model of Ru-1212 as shown in Fig. 12(c), that is, Ru-1212 is composed of superlattice domains of about 10 nm in width distinguished by the P and I superlattices along the c direction.

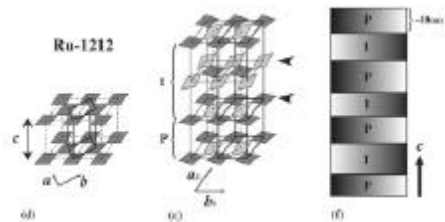

**Fig. 13**. An example of a proposed domain-structure model for Ru-1212; (d) the fundamental lattice of Ru-1212; (e) the two possible superlattices (P and I superlattices) of Ru-1212; and (f) an example of a proposed domain-structure model for Ru-1212.

From microstructures and superstructures studies of the Ru-1222 and Ru-1212 magneto-superconductors, it is revealed that both Ru-1222 and Ru-1212 are composed of nm-size domains stacked along the c direction. Origin of such domain is the ordering of rotated $RuO_6$ octahedral around the c-axis, which leads to the formations of characteristic superstructures. In Ru-1212, two types of superstructures with primitive (P) and body-centered (I) symmetries are derived, and they alternately repeated with nm-size periodicity along the c-axis. It is of great interest that Ru-1212 consists of domains of two crystallographically different superstructures, while Ru-1222 consists of domains of the single crystallographically identical superstructure, with base-centered orthorhombic symmetry, mutually related by 90° rotation around the c-axis.

### 3.5. Magneto-superconductivity and magnetic characteristics: SQUID results

#### 3.5.a. Non-superconducting Ru-1212 samples

Figure 14 depicts both ZFC (zero-field-cooled) and FC (field-cooled) magnetic moment ($M$) *versus* temperature ($T$) plots for various $RuSr_2GdCu_2O_{8-\delta}$ samples with an applied field of 10 Oe [43]. As is seen from this figure the ZFC and FC magnetisation curves show a significant branching around 145 K for the as-synthesized sample with a further sharp drop in magnetisation around 2.6 K.

The branching of ZFC and FC curves at 145 K originates from the magnetic ordering of Ru moments and the sharp peak at 2.6 K is due to antiferromagnetic ordering of Gd moments. Interestingly in presently studied non-superconducting $RuSr_2GdCu_2O_{8-\delta}$ samples the ordering temperature of Ru moments appears to be 13 K higher than the value reported for similar but superconducting samples [8,15,33]. Both the ZFC and FC branching about 145 K and the sharp peak at 2.6 K are seen for all the samples including the high-$O_2$-pressure annealed and the argon annealed ones. None of the samples show any traces of superconductivity down to 2 K, even with very low field measurements at 1.5 Oe. Worth reminding is the fact that these samples of ours contain no traces of $SrRuO_3$, (see Fig. 15). These non-superconducting $RuSr_2GdCu_2O_{8-\delta}$ samples showed a similar ferromagnetic component at 5 K, as for reported superconducting samples.

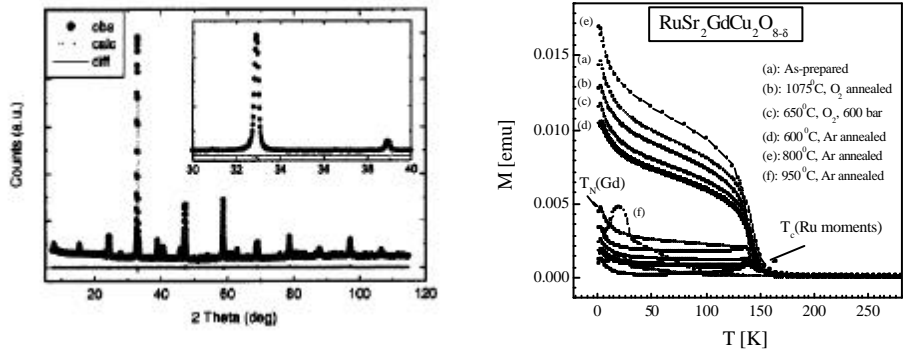

**Figs. 14&15.** *M-T* plots & XRD for various non-superconducting RuSr$_2$GdCu$_2$O$_{8-\delta}$ samples.

Except the fact that the present samples were non-superconducting and their phase purity was comparatively better than for reported samples, no other visible difference is observed, though it seems that non-superconducting samples have a bit higher magnetic ordering temperature. Worth emphasising is the fact that even after various post-annealing steps, the superconductivity could not be achieved. Also observed is the fact that the magnetic ordering temperatures of Ru or Gd moments were not affected by the post-annealing steps. This highlights the fact, discussed in section 3.2, that there is not much room left for tuning the oxygen stoichiometry of Ru-1212.

The magnetic susceptibility ($\chi$) of the compound follows the paramagnetic behaviour above the magnetic ordering temperature of Ru moments. Considering Gd to be in trivalent state with a localised moment of 8 $\mu_B$ (same as in Gd-based Cu-1212), the calculated moment from Curie-Weiss relation for Ru in paramagnetic state is around 1.0 $\mu_B$, suggesting Ru to be in pentavalent state in Ru-1212. However we should like to mention that moment extraction from Curie-Weiss relation can not be conclusive, without properly considering the exact state of Cu and the effect of possible crystal fields on the magnetic susceptibility of the compound. In fact effective paramagnetic moment for Ru in Ru-1212 has been reported as high as 3.17(4) $\mu_B$ based on high temperature (up to 900 K) fitting of the magnetic susceptibility [17]. The valence of Ru extracted from the magnetic susceptibility data vary from one report to another. We believe the fitting of high temperature magnetic susceptibility to simple Curie-Weiss relation is futile in determining the valence of Ru in rutheno-cuprates like Ru-1212 and Ru-1222, but spectroscopic methods such as XANES are more conclusive in determining the valence state of Ru.

### 3.5.b Magneto-superconductivity of as-synthesized Gd/Ru-1212 samples

Figure 16 shows the $\chi$-*T* behaviour in the temperature range of 5 K - 160 K for another as-synthesized RuSr$_2$GdCu$_2$O$_{8-\delta}$ sample with an applied field of 5 Oe, in both ZFC and FC situations. The ZFC and FC curves start branching around 140 K with a cusp at 135 K and a diamagnetic transition around 20 K in the ZFC part. The down-turn cusp at 135 K is indicative of antiferromagnetic nature of Ru-spin ordering. Interestingly for the same sample annealed in 100-atm O$_2$ atmosphere the diamagnetic transition was not observed down to 5 K (curve not shown) [43].

For the as-synthesized sample the FC part is seen increasing and later saturating probably due to contribution from paramagnetic Gd moments. Inset of Fig. 16 shows the isothermal *M versus* applied field (*H*) behaviour for this sample. The isothermal magnetization as a function of magnetic field may be viewed as:

$$M(H) = \chi H + \sigma_s(H), \qquad (1)$$

where $\chi H$ is the linear contribution from antiferromagnetic Ru spins and paramagnetic Gd spins and $\sigma_s(H)$ represents the weak ferromagnetic component of the Ru sublattice. The contribution from the weak ferromagnetic component starts to appear only below 100 K and at higher fields above 3 T. Above this temperature the *M-H* plot remains purely linear.

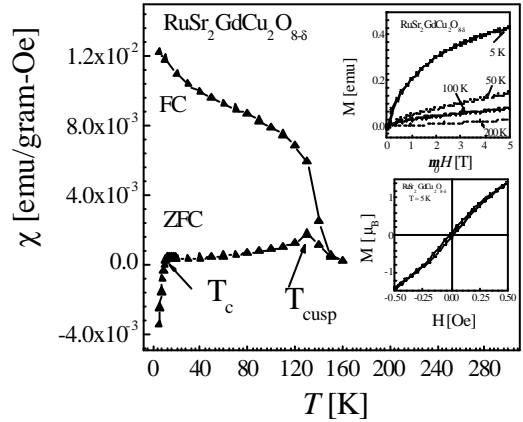

**Figs. 16.** $\chi$-$T$ plot for an as-synthesized, superconducting Gd/Ru-1212, the insets show the $M$-$H$ loops for the same.

Appearance of ferromagnetic component at low $T$ within antiferromagnetically ordered Ru spins can happen due to slight canting of spins. Published neutron diffraction data clearly indicate such a possibility [11,12]. Non-linearity in $M$-$H$ appears at high fields above 3 T. The $M$-$H$ loop for the sample is shown in the lower inset of Fig. 16.

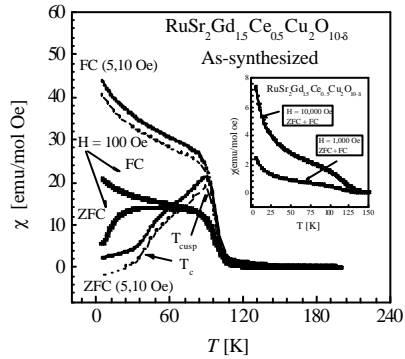

**Figs. 17.** $\chi$-$T$ plot for an as-synthesized, superconducting Gd/Ru-1222 samples.

### 3.5.c. Magneto-superconductivity of as–synthesized Gd/Ru-1222 samples

Figure 17 shows the magnetic susceptibility $\chi$-$T$ behaviour in the temperature range of 5 to 160 K for an as-synthesized $RuSr_2(Gd_{0.75}Ce_{0.25})_2Cu_2O_{10-\delta}$ sample under applied fields of 5, 10 and 100 Oe, measured in both ZFC and FC modes. In an applied field of 5 Oe, the ZFC and FC curves start branching at around 140 K with a sharp upward turn at 100 K. The ZFC branch shows further a cusp at 85 K and a diamagnetic transition around 30 K. This is in general agreement with other reports [7,27,28]. The ZFC curve does not show any diamagnetic transition in applied fields of 10 and 100 Oe, but the transition is marked with a change in the slope of the ZFC curves. As the field strength exceeds a certain threshold value the positive contribution from both Gd and Ru moments overcomes the negative contribution from superconductivity to the magnetic susceptibility. Interestingly the ZFC - FC branching temperature of 140 K in 5 Oe field decreases to around 60 K in an applied field of 100 Oe. This can be considered as a weak ferromagnetic behaviour. In fact no ZFC - FC branching is observed down to 5 K in 1,000 and 10,000 Oe fields where both the anomaly and the irreversibility in ZFC and FC branches look to be washed out, see inset in Fig. 17. The down-turn cusp at 85 K in low fields is indicative of antiferromagnetic or spin-glass nature of Ru spins. The FC curve is seen increasing or saturating due to the contribution from paramagnetic Gd spins.

**3.5.d. Magneto-superconductivity of "100-atm O$_2$-annealed" Gd/Ru-1222 samples**

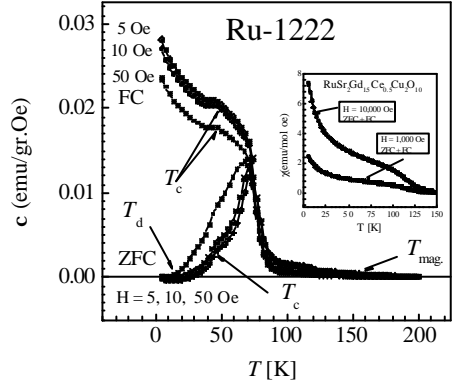

**Fig. 18.** χ-$T$ plots for a 100-atm O$_2$-annealed Gd/Ru-1222 sample

Fig. 18 shows the magnetic susceptibility (χ) vs. $T$ behaviour in the temperature range of 5 to 200 K for "100-atm O$_2$-annealed" Ru-1222 sample under applied fields of 5, 10 and 50 Oe, measured in both zero-field-cooled (ZFC) and field-cooled (FC) modes. In an applied field of 5 Oe, the χ vs. $T$ show the branching of zero-field-cooled (ZFC) and field-cooled (FC) curves at around 90 K ($T_{irr}$), a step like structure in both at around 40 K ($T_c$) and further a diamagnetic transition around 40 K ($T_d$) in the ZFC magnetization.

Though the ZFC and FC magnetization branching is seen at around 90 K, the magnetic behaviour starts deviating from normal paramagnetic relation at much higher $T$ say 160 K. The characteristic temperatures $T_{irr}$, $T_c$, and $T_{mag}$ are weakly dependent on $H <$ 100 Oe. For higher $H >$ 100 Oe, both ZFC and FC are merged with each other, and only $T_{mag}$ could be seen, see inset Fig. 16. This is in general agreement with earlier reports [7,27,28]. In fact no ZFC - FC branching is observed down to 5 K in both 1,000 and 10,000 Oe fields and both the anomaly and the irreversibility in both ZFC and FC branches look to be washed out. The ZFC curve did not show any diamagnetic transition ($T_d$) in $H >$ 50 Oe. The magnetization data at $H$ = 10 Oe, show nearly the same characteristics as for $H$ = 5 Oe. A low field (-100 Oe = $H$ = 100 Oe) $M$ vs. $H$ loop for currently studied Ru-1222 compound is shown in Fig.19. Interestingly the diamagnetic signal starts decreasing above applied fields of 25 Oe, and turns to zero at say 40 Oe. The compound seems to have a lower critical field ($H_{c1}$) of around 25 Oe. Interestingly the $M$ vs. $H$ plot shown in Fig. 19 does not appear to be a normal HTSC case. We will discus the low field (-100 Oe = $H$ = 100 Oe) $M$ vs. $H$ loop of Fig.19 again after further magnetic characterization in next section.

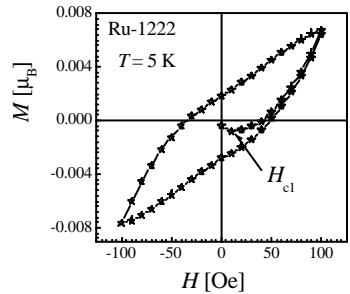

**Fig. 19.** ($M$) vs. applied field ($H$) plots for a 100-atm O$_2$-annealed Gd/Ru-1222 sample

To elucidate the magnetic property of Ru-1222 we show isothermal magnetization ($M$) vs. applied field ($H$) behaviour at various $T$ (Fig. 20). Clear $M$ vs. $H$ loops are seen at 5, 10, 20, and 40 K. The applied fields are in the range of -2000 Oe = $H$ = 2000. At 5 K, the returning moment ($M_{rem}$) i.e. the value of magnetization at zero returning field and the coercive filed ($H_c$) i.e. the value of applied returning field to get zero magnetization are respectively 0.35 μ$_B$ and 250 Oe. Worth mentioning is the fact that Gd (magnetic rare earth) in the compound orders magnetically below 2 K and Ce is known to be in tetravalent non-magnetic state hence the $M_{rem}$ and $H_c$ arising from the ferromagnetic hysteresis loops do belong to Ru only. Hysteresis loops are not seen for $M$ vs. $H$ plots above 80 K. For various hysteresis loops being observed from $M$ vs. $H$ plots below 80 K, the values of both $M_{rem}$ and $H_c$ decrease with $T$. The plots for both are shown in upper and lower insets of Fig.20. Both $M_{rem}$ and $H_c$ of 0.35 $\mu_B$ and 250 Oe being observed for Ru-1222 are much higher than reported for other magneto-superconductor Ru-1212 [8,16]. For Ru-1212 the hysteresis loops are reported quite narrow with $M_{rem}$ and $H_c$ of 0.085 $\mu_B$ and 10 Oe respectively. This indicates that in Ru-1222 the ferromagnetic domains are less anisotropic and more rigid.

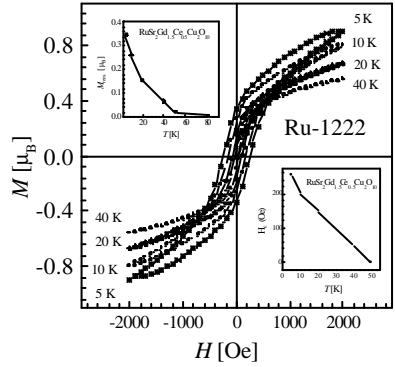

**Fig. 20.** $M$ - $H$ loops for a 100-atm $O_2$-annealed Gd/Ru-1222 at 5, 10, 20 and 40 K with -2000 = $H$ = 2000 Oe. The upper and lower insets of the figure show $M_{rem.}$ vs. $T$ and $H_c$ vs. $T$ plots for the same.

The isothermal magnetization as a function of magnetic field at 5 K with higher applied fields; 70000 Oe = $H$ = 70000 Oe is shown in Fig.21. The saturation of the isothermal moment appears to occur above say 5 T applied fields. The contribution from the ferromagnetic component starts to appear below 100 K. The presence of the ferromagnetic component is confirmed by hysteresis loops being observed at 5, 10, 20 and 40 K in the $M$ vs. $H$ plots, (see Fig. 19). Ru spins order magnetically above say 100 K with a ferromagnetic component within ($M_{rem}$, $H_c$ = 0.35 $\mu_B$, 250 Oe) at 5 K. As far the value of higher field (> 5 T) saturation moment is concerned, one can not without ambiguity extract the value for Ru contribution. Basically besides paramagnetic Gd contribution at 5 K, the contribution from Cu can not be ignored, which in an under-doped HTSC compound contributes an unknown paramagnetic signal to the system. For paramagnetic Gd contribution the theoretical plot at 5 K is shown in the inset of Fig. 21. After taking out the Gd contribution from Ru-1222 effective moment in Fig.6, a value of ~ 0.75 $\mu_B$ is obtained for effective near saturation moment of Ru. This value is lees than for $Ru^{5+}$ low spin state ordering. In Gd/Ru-1212 compound, based on various magnetization data the $Ru^{5+}$ state is reported with an effective saturation moment of nearly 1$\mu_B$ [8,11], which ironically differs with more recent magnetic analysis [17].

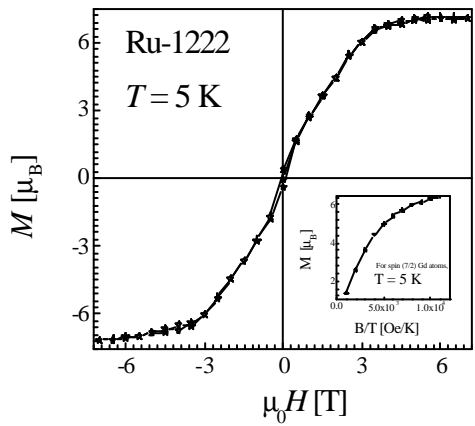

**Fig. 21** $M$ - $H$ plot for 100-atm $O_2$-annealed Gd/Ru-1222 sample at $T$ = 5 K, the applied field are in the range of –70000 Oe = H = 70000 Oe. The inset shows the theoretical plot for paramagnetic Gd contribution to the system.

Superconductivity is seen in terms of diamagnetic transition at below $T_d$, and $T_c$ ($R$=0) at slightly higher temperature. It is known earlier that due to internal magnetic field, these compounds are in a spontaneous vortex phase (SVP) even in zero external field [44]. For $T_d < T < T_c$ the compound remains in mixed state. Hence though $R$= 0 is achieved at relatively higher temperatures the diamagnetic response is seen at much lower $T$ and that also in quite small applied magnetic ($H_{c1} < 25$ Oe) fields. Now we can understand the $M$ vs. $H$ loop being shown in Fig.17. As discussed in previous section clear ferromagnetic component is seen in the compound at 5 K. Hence at 5K both ferromagnetic and the superconducting hysteresis loops are present in the $M$ vs. $H$ magnetization data, and at low applied fields viz. -100 Oe = H = 100 Oe, the compound simply exhibit the superimposition of the both, which is the case in Fig.19.

### 3.5.e. Magneto-superconductivity of "N$_2$-annealed" Gd/Ru-1222 samples

The $\chi$-T behaviour in the temperature range of 2 to 300 K for N$_2$-annealed Ru-1222 sample in an applied fields of 100 Oe, measured in both zero-field-cooled (ZFC) and field-cooled (FC) modes, is shown in Fig.22. The general shape of FC and ZFC magnetization plots is similar to that for earlier discussed samples. The only interesting change is that $T_{mag}$ (defined earlier) has increased to 106 K for N$_2$-annealed sample. Worth mentioning is the fact that N$_2$-annealed sample is not superconducting down to 2 K.

Figure 23 depicts the $M$-$H$ plot at 5 K for N$_2$-annealed sample. This plot is similar to that observed for earlier discussed samples. The zoomed ferromagnetic component is shown in the inset. The interesting difference, when compared with as-synthesized and 100-atm-O$_2$ annealed samples, is that $M$ does not saturate in applied fields of up to 7 Tesla. This is in contrast to the $M$-$H$ plot for air/100-atm O$_2$-annealed samples at 5 K for which M saturates in a field of 6 Tesla (see Fig. 19). The $M$-$H$ plot for N$_2$-annealed sample is further zoomed in applied field of –900 Oe to 900 Oe, and shown in the inset of Fig.21. Ferromagnetic loop is seen clearly with $M_{rem}$ (2emu/gram) and $H_c$(170Oe). Relatively lower characterestic values of $M_{rem}$ and $H_c$ for N$_2$-annealed sample can be discussed on the basis of SAED and HRTEM studies. Our detailed micro-structural studies earlier for Ru-1222 showed that the observed super-lattice structures due to tilt of RuO$_6$ octahedra [40] might be coupled with the weak ferromagnetic domains constructed by ordering of the canted Ru moments below the magnetic transition temperature ($T_{mag}$). Hence ferromagnetic domains coupling depends on the long range ordering of tilted RuO$_6$ octaherdas in a given Ru-1222 system. In N$_2$-annealed sample, the long-range superstructures may break down relatively at smaller length scale than for other samples due to less oxygen in RuO$_6$ octahedra of the same giving rise to weak coupling of the ferromagnetic domains. This will give rise to lower values of $M_{rem}$ and $H_c$. This also explains the fifth point regarding the observed saturation of $M$-$H$ curve for other samples and not for N$_2$-annealed sample. The saturation of $M$-$H$ is dependent on the long range coupling of aligned ferromagnetic domains, which is observed for air-annealed sample only. Long range coupling of aligned moments is directly dependent on the stability of RuO$_6$ octahedra tilt angle superstructures, which is certainly less for N$_2$-annealed sample due to break down in homogenous oxygen content close to 6.0 in the octahedra.

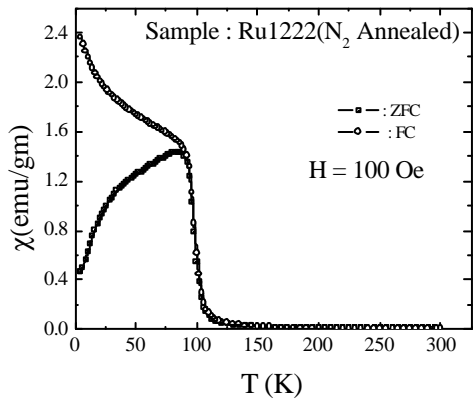

**Fig. 22.** Magnetic susceptibility ($\chi$) vs. temperature (T) plot for N$_2$-annealed Ru-1222 in both ZFC and FC modes with applied field of 100 Oe.

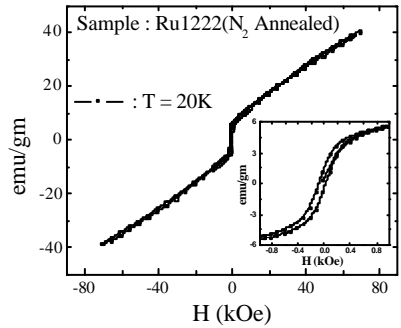

**Fig. 23.** Magnetization-Field ($M$–$H$) hysteresis loop for the $N_2$– annealed Ru-1222 at 5 K, inset shows the same for –900Oe = H = +900Oe.

### 3.5.f. Magneto-superconductivity of HPHT synthesized Y/Ru-1212 sample

One of the puzzles in understanding the magnetization data of Gd/Ru-1212, is not only the antagonising nature of superconducting and magnetic order parameters, but also the presence of magnetic Gd (8µB) along with the possibly magnetic Cu moment (Cu is paramagnetic in under-doped HTSC), which hinders in knowing the exact Ru spins magnetic contribution to the system. Ru-1212 can be formed with non-magnetic RE (rare earth) Y instead of Gd, but only with HPHT (high pressure high temperature) synthesis technique. Magneto-superconductivity of Y/Ru-1212 is discussed below;

Figure 24 shows both ZFC (zero-field-cooled) and FC (field-cooled) magnetic susceptibility versus temperature ($\chi$ vs. $T$) plots for $Ru_{0.9}Sr_2YCu_{2.1}O_{7.9}$ sample, in various applied fields of 50, 70, 100, 300, and 1000 Oe. As seen from this figure the ZFC and FC magnetization curves show a significant branching at around 145 K. The branching of ZFC and FC at this temperature is indicative of the magnetic ordering of Ru moments. It is known from neutron diffraction studies that Ru moments order antiferromagnetically at around 133 K for Gd/Ru-1212 compound [11,12]. As the real nature of the magnetic ordering of Ru moments is still debated, we denote this temperature as $T_{mag}$. In an earlier report on HPHT synthesized Y/Ru-1212 compound $T_{mag}$ of around 150 K was observed, which is in close agreement to the current value. With an increase in applied field (10 Oe > $H$ > 1000 Oe ) basically no change is observed in $T_{mag}$. The ZFC part of magnetic susceptibility at low $T$ below 30 K, shows clear diamagnetic transitions till applied fields of 300 Oe. The extent of diamagnetic signal is field dependent, and is not observed at 1000 Oe. The diamagnetic signal onset temperature is described as superconducting transition temperature ($T_c$). Worth noting is the fact that the diamagnetic signal being observed in ZFC measurements does not saturate down to 5 K.

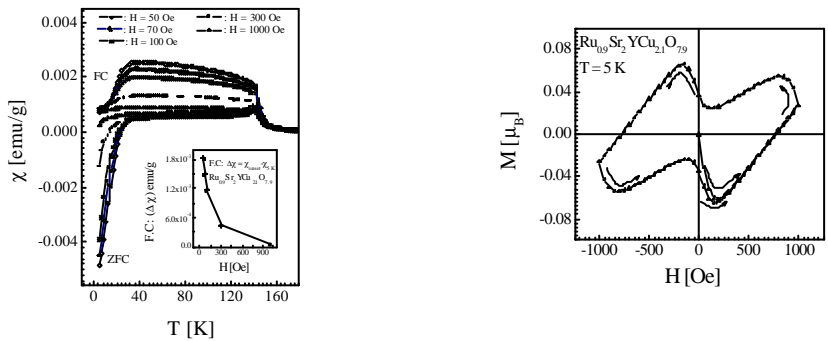

**Fig.24&25.** Magnetic susceptibility versus temperature ($\chi$ vs. $T$) plots for $Ru_{0.9}Sr_2YCu_{2.1}O_{7.9}$ sample, in various applied fields of 50, 70, 100, 300, and 1000 Oe, inset shows the difference of magnetic susceptibility ($\Delta\chi$) between $T_c$ onset and 5 K for field-cooled (FC) transition for various fields, & the M vs H plot for the same at low fields at 5 K.

On the other hand, the FC part of the magnetic susceptibility remains positive down to 5 K. The rise in the FC part of magnetic susceptibility is nearly saturated below around 50 K, indicating towards a ferromagnetic transition, followed by a dip below 30 K with further saturation at temperatures lower than 10 K. The dip is FC susceptibility below 30 K, is dependent on applied field, higher is the field and less is the dip. For applied magnetic field of 1000 Oe, no dip is observed in magnetic susceptibility down to 5 K. Interestingly the dip in FC transitions occurs exactly at the temperature being defined as $T_c$, in ZFC transition. Though in FC transition the diamagnetic signal is not observed, the observation of clear dip at $T_c$ guarantees the observation of bulk nature of superconductivity in the compound. This approves the bulk nature of superconductivity in the compound. The nature of bulk superconductivity was evidenced earlier in Gd/Ru-1212 compound by dip in FC transition at $T_c$ at very low applied fields (< 2.5 Oe) [8]. In our Y/Ru-1212 compound bulk superconductivity is evidenced till applied fields of 300 Oe. The dip in FC magnetization at $T_c$ decreases with the applied field. Numerically the dip in FC magnetization at $T_c$ is defined as the difference of magnetic susceptibility ($\Delta\chi$) between $T_c$ onset and 5 K, which is plotted in inset of Figure 24. It is clear from the inset in Figure 24, that $\Delta\chi$ decreases with an increase in $H$. However clear $\Delta\chi$ values are seen up to $H$ = 300 Oe, confirming the presence of bulk superconductivity in the compound at least till these applied fields. Estimated Meissner superconducting volume fraction (calculated from $\Delta\chi$ value) is nearly 15 % at $H$ = 50 Oe and above 4 % at $H$ = 300 Oe. Worth discussing is the fact that the values obtained above are for superconductivity which is under internal magnetic field from Ru sub-lattice ferromagnetic field along with the external applied field. Ru-1212 and Ru-1222 compounds are supposed to be in spontaneous vortex phase (SVP) even in zero applied fields due to the application of internal magnetic field [8,44]. It is only with presently studied HPHT synthesized Y/Ru-1212 compound that sufficient dip in terms of $\Delta\chi$ is seen in FC transition up to $H$ = 300 Oe, otherwise in earlier reports on Gd/Ru-1212 the FC transition is not observed even in quite low fields of $H$ < 10 Oe.

Worth noting is the fact, that though the dip in FC magnetic susceptibility is saturated below say 10 K, the ZFC transition below $T_c$ is not saturated down to 5 K. It seems that though the nature of superconductivity being observed at $T_c$ is of bulk nature as indicated by dip and further saturation in FC magnetization, the same is not well connected and hence missing the most of surface screening currents. This is the reason that the diamagnetic transitions seen in ZFC part of magnetic susceptibility are not saturated down to 5 K. Possibly the superconducting domains of bulk nature are being disconnected with non-superconducting clusters. This results in a two-phase (bulk-superconducting/non-superconducting clusters) system existing in the compound. This gets further credence from the fact, that no $T_c$ (R = 0) state is observed in the compound. Only a partial drop in resistance is observed at around 30 K. There is a strong possibility of the formation of *SIS* (Superconducting-Insulating-Supercoducting) or *SNS* (Superconducting-normal-superconducting) junctions in the compound. Depending on the width ($d$) of *I* or *N* blocks between superconducting clusters and the coherence length (?) of the superconductor, one may or may not get the R = 0 state. In present situation it seems that $d \gg$ ?. Figure 25 depicts the *M* vs. *H* plot for the presently studied $Ru_{0.9}Sr_2YCu_{2.1}O_{7.9}$ compound. The applied field $H$ is in the range of 0 < $H$ < 1000 Oe. It is seen from the figure that the magnetization is initially increasingly negative up to say 50 Oe, which later moves towards lower negative values and finally turns to positive above 700 Oe. The sequence of the *M-H* loop is though similar to a normal Type II superconductor, the shape of the curve is rather complicated. A closer look at the present *M-H* loop rather indicates towards the superimposition of a superconducting (negative diamagnetic) and the ferromagnetic (positive) magnetization. The ferromagnetic component of Ru-spins magnetic ordering being present in the compound at 5 K is riding over the superconducting signal. For example at returning fields from 1000 Oe, first a peak is observed in *M-H* loop and later at lower fields of < 300 Oe a clear dip is observed in the loop, before further increase at increasing negative fields. At returning lower fields the diamagnetic signal due to superconductivity again dominates the positive contributions from ferromagnetic component and hence the dip is observed in the loop. Calculating the critical current density ($J_c$) from magnetic loop is not feasible, because the exact contribution of the ferromagnetic Ru spins is not known. Also the nature of interactions between the two order parameters (superconducting and magnetic) is not understood. The simple summation of the two signals might not be the exact situation. Further as we will show in the next section that the ferromagnetic contribution of Ru-spins magnetic ordering at 5 K is not yet clear. To our knowledge, ours is the first sample where an *M-H* loop is obtained which is being characteristic of the superconducting and ferromagnetic signals being riding over each other at least until few hundred Oe applied fields. Magnetization is a bulk measurement technique and hence it is no guarantee for the co-existence of superconductivity and ferromagnetism in the material at the microscopic level within the same phase.

In Figure 26 are shown various *M-H* loop for the compound at various *T* of 5, 20, 50, 100, 120 and 150 K, in applied fields of -70000 < *H* < 70000 Oe. Depicted *M-H* loops clearly show the ferromagnetic like behaviour at least below 120 K. In fact even at 150 K the *M-H* loop is not completely linear, indicating the fact that some ordered magnetic domains do exist even at this *T* also. The neutron diffraction studies on Gd/Ru-1212 compound earlier concluded that Ru-spins order antiferromagnetically at high temperatures and the ferromagnetic component is developed due to canting of moments only say below 20 K [11,12].

The *M-H* loops for our currently studied Y/Ru-1212 compound clearly demonstrate towards the ferromagnetic order with the magnetization getting nearly saturated below 20 K. As far as the saturation moment values etc are concerned one can observe that though the complete saturation of moments is not achieved at applied fields of as high as 7 T, the near saturation value observed at 7 T and 5 K is 1.17μB, which is nearly the same at 20 K also. Interestingly this value is higher than as expected theoretically for magnetic ordering of low spin (1/2) states of

$Ru^{5+}$ and considerable less than for high spin (3/2) state. In such a situation we rather believe a mixed valence state of Ru ($Ru^{5+}/Ru^{4+}$) would be more appropriate.

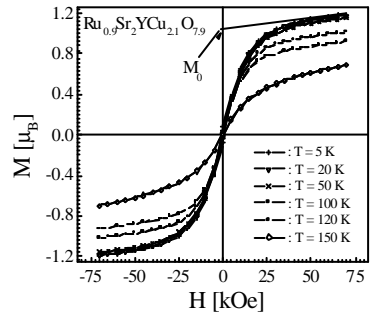

**Fig. 26**. M vs. H plot for the $Ru_{0.9}Sr_2YCu_{2.1}O_{7.9}$ compound at $T$ = 5, 20, 50, 100, 120 and 150 K, the applied field H are in the range of -70 kOe < H < 70 kOe.

### 3.5.g. Magnetism of HPHT synthesized Ln/Ru-1222 sample

Figure 27 show both zero-field-cooled (zfc) and fc magnetic susceptibility versus temperature ($\chi$ vs. $T$) plots for the Y/Ru-1222 sample, in external fields of 5 and 20 Oe. As seen from this figure the fc magnetization curve shows an increase near 150 K, followed by a significant jump at around 100 K. The zfc branch shows a rise in magnetization at around 110 K and a cusp like down turn in magnetization at 100 K. In general the magnetization behaviour of the compound can be assigned to a weak ferromagnetic transition at around 100 K. However what is not understood is the initial rise of fc magnetization at 150 K. The interesting difference is that in HPHT synthesized Ln/Ru-1222 compounds the 150 K transition in fc magnetization is more pronounced than for reported Gd/Ru-1222 [7,27,34]. Figure 28 shows both zero-field-cooled (zfc) and fc magnetic susceptibility versus temperature ($c$ vs. $T$) plots for the Ln/Ru-1222 samples, with Ln = Ho, Dy. The general behaviour of the all the samples is similar to that as for Y/Ru-1222. The fc transition is seen in both the samples at 150 K. The zfc cusp and the diamagnetic transition are though Ln dependent, but essentially in the same temperature ranges.

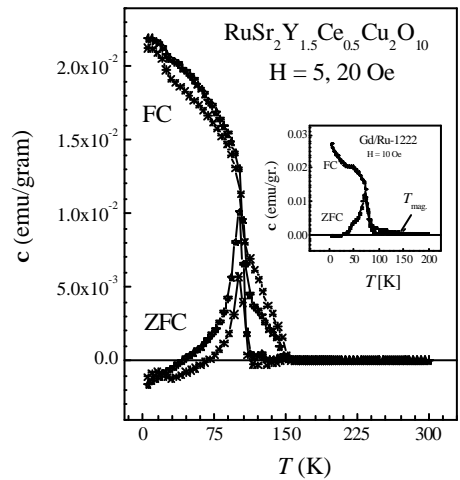

**Fig.27.** Magnetic susceptibility versus temperature ($\chi$ vs. $T$) plots for Y/Ru-1222 sample, in various applied fields of 5, and 20 Oe, inset shows the same for reported Gd/Ru-1222 compound.

Though the studied samples are almost single phase in x-ray, the minute impurities like $SrRuO_3$ or Ln/Ru-1212 might be responsible for the fc transition at 150 K. To exclude such a possibility we would like to stress that in Ln/Ru-1212 compounds the 150 K fc transition is followed by a cusp in zfc at same temperature and is also Ln dependent. For Ln = Ho and Dy in Ln/Ru-1212 the Ru spins magnetic ordering temperature is reported to be at 170 K [18]. In Ln/Ru-1222 compounds we do not observe a cusp in zfc and also the FC transition at 150 K is not Ln dependent.

Hence the possible origin of fc transition at 150 K due to Ln/Ru-1212 is excluded. At this juncture we believe that the 150 K transition in fc magnetization of Ln/Ru-1222 compounds is intrinsic to this phase. This gets credence from the fact, that though Ln = Y sample unlike others contains small impurity of $SrRuO_3$, the 150 K transition in fc is same for all the studied compounds. In widely studied Gd/Ru-1222 compound also the rise in FC magnetization is reported at around 160-180 K, and was associated with an antiferromagnetic transition of Ru spins [7,27,34].

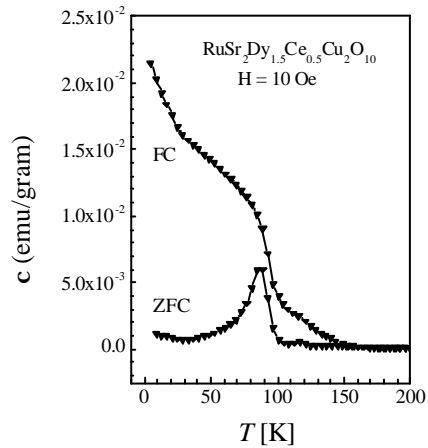

**Fig. 28.** ($\chi$ vs. $T$) plots for Dy/Ru-1222.

We did couple of $M$ vs $H$ experiments for Ln = Ho sample at various temperatures of 130, 120, 110, 100 and 5 K, the results are shown in Fig.29 (a, b, c, d and e). At temperature of 130 and 120 K the $M$ vs. $H$ hysteresis loops exhibit an antiferromagnetic like structure with canted moments, though at 110 K the same possess more like an S-type spin-glass shape. It seems that the re-orientation of Ru-spins or change in canting angle takes place at 110 K. Further at 5 K, it is more like a ferromagnetic loop. The 5 K, data for non-magnetic Ln = Y will be discussed in Fig. 28.

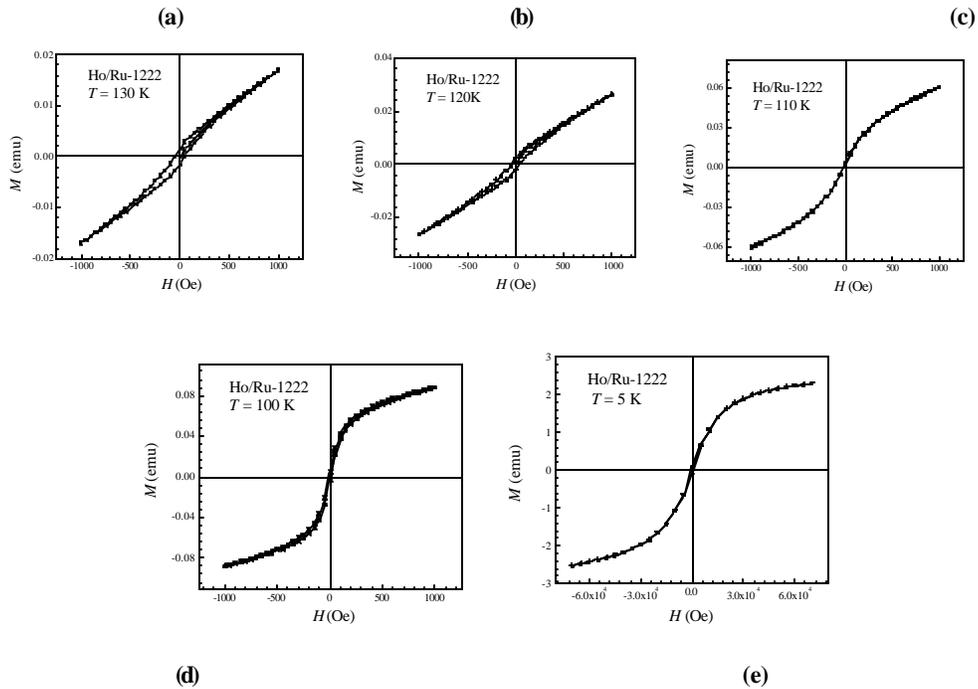

**Fig.29**. $M$ vs. $H$ plot for the Ho/Ru-1222 compound at $T =$ (a) 130 K, (b) 120 K, (c) 110 K, (d) 100 K and (e) 5 K.

One wild speculation might be that before 100 K weak ferromagnetic transition, the Ru spins go through a spin-glass like transition at around 110 K and an antiferromagnetic transition at even higher temperature of say 150 K. Without detailed magnetic structure from neutron scattering experiments, it is difficult to comment on exact nature of the magnetism of various Ru-1222 compounds. Ironically, as we mentioned in the introduction, yet no detailed magnetic structure refinements from neutron scattering experiments are available for Ru-1222 compounds. Our current results are one step a head to widely reported Gd/Ru-1222 compound that the 150-160 K transition in magnetization before weak ferromagnetism at 110 K can not be left unnoticed as the same is quite sharp in our samples and is universal to all studied Ln/Ru-1222 compounds.

The zfc and fc significant branching temperature of 100 K for Y/Ru-1222 is relatively higher than previously reported ~ 80 K for Gd/Ru-1222. For reference, reported [34] $\chi$ vs. $T$ plot for Gd/Ru-1222 is shown in inset of Fig.26. Interestingly for magnetic ordering temperature for Gd/Ru-1212 of ~133 K was also found to be relatively lower than for HPHT synthesized Y/Ru-1212 (~150 K) [12,16,18]. The zfc part of magnetic susceptibility at low temperature below 70 K shows a clear shoulder with further weak diamagnetic transition below ~50 K. The zfc curve did not show any diamagnetic transition ($T_d$) in $H$ = 100 Oe. The shoulder at 70 K is known as $T_c$ (superconducting transition temperature) from various experiments in Gd/Ru-1222. It is known earlier that due to internal magnetic field, these compounds are in a spontaneous vortex phase (SVP) even in zero external field [44]. For $T_d < T < T_c$ the compound remains in mixed state. Hence though possibly superconductivity is achieved at relatively higher temperatures the diamagnetic response is seen at much lower $T$ and that also in quite small applied magnetic fields [7,27,34]. Worth mentioning is the fact that the electrical transport measurements being necessary for confirmation of superconductivity are yet not carried out on presently synthesised Ln/Ru-1222 compounds. Hence the superconductivity as such cannot be confirmed, detailed various measurements are underway and will be reported shortly.

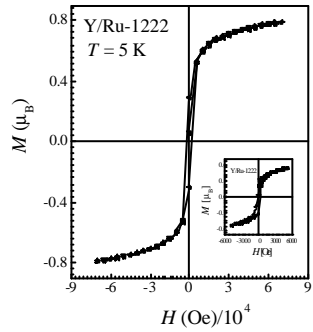

**Fig. 30** $M$ vs. $H$ plot for the Y/Ru-1222 compound at $T = 5$ K, the applied field H are in the range Of -70 kOe < $H$ < 70 kOe. The inset shows the $M$ vs. $H$ plot for the same at 5 K, with -50000 = H = 5000 Oe.

The isothermal magnetization as a function of magnetic field at 5 K with higher applied fields; 70000 Oe = $H$ = 70000 Oe for Ln = Y sample is shown in Fig.30. The saturation of the isothermal moment appears to occur above say 4 T applied fields. The presence of the ferromagnetic component is confirmed by hysteresis loops being observed at 5 K in the $M$ vs. $H$ plots, (see inset Fig. 30). Ru spins order magnetically above say 100 K with a ferromagnetic component within ($M_{rem}$, $H_c$ = 0.30 $\mu_B$, 150 Oe) at 5 K. As far the value of higher field (> 4 T) saturation moment is concerned, one cannot without ambiguity extract the value for Ru contribution, because the contribution from Cu cannot be ignored. In an under-doped HTSC compound Cu contributes an unknown paramagnetic signal to the system. Without considering the Cu contribution an effective moment of ~ 0.80 $\mu_B$ is obtained for Ru. This value is lees than for $Ru^{5+}$ low spin state ordering.

### 3.5.h. Magnetism of HPHT synthesized Ln/Ru-1232 sample

Figure 31 show both zero-field-cooled (zfc) and fc magnetic susceptibility versus temperature ($\chi$ vs. T) plots for the Y/Ru-1232 sample, in external fields of 10 Oe. As seen from this figure the fc magnetization curve shows an increase near 90 K ($T_{mag.}$). Furtrher, the zfc branch shows a cusp like down turn in magnetization at around 70 K. In general the magnetization behaviour of the compound can be assigned to a weak ferromagnetic transition at around 90 K. The zfc branch also shows a step like structure at around 45 K ($T_c$) and a diamagnetic transition around 35 K ($T_d$).

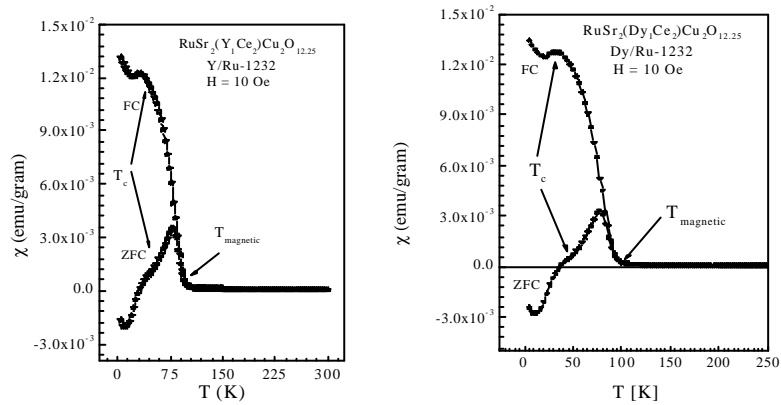

**Figs. 31&32.** Magnetic susceptibility versus temperature ($\chi$ vs. *T*) plots for Y/Ru-1232 & Dy/Ru-1232 samples, in various applied fields of 5 Oe.

Figure 32 shows both zero-field-cooled (zfc) and fc magnetic susceptibility versus temperature ($\chi$ vs. T) plots for the Dy/Ru-1232 sample. The general behaviour is similar to that as for Y/Ru-1232. The fc transition is seen at 90 K. The zfc cusp at around 70 K, the step like structure ($T_c$) at around 40 K and diamagnetic transition ($T_d$) is seen at around 35 K for Dy/Ru-1232 compound. Superconductivity is seen in terms of diamagnetic transition below $T_d$. It is known earlier that due to internal magnetic field, these compounds are in a spontaneous vortex phase (SVP) even in zero external field. For $T_d < T < T_c$ the compound remains in mixed state. Hence though superconductivity might be achieved at relatively higher temperatures the diamagnetic response is seen at much lower T and that also in quite small applied magnetic fields.

To further elucidate on the magnetization of these compounds, the isothermal magnetization as a function of magnetic field at 5, 20, 50, 80 and 120 K with applied fields; -70000 Oe = H = 70000 Oe for Ln = Dy sample is shown in Fig.33. The saturation of the isothermal moment appears to occur above say 4 Tesla applied fields at 5 K. Further increase in magnetization above say 4 Tesla is due to the contribution from Dy moments. At higher temperatures of 20, 50, 80, 100 and 120 K the near saturation of M vs. H is not seen. The presence of the ferromagnetic component is confirmed by hysteresis loop being observed at 5 K in the M vs. H plots (-1000 Oe = H = 1000), (see inset Fig. 33). Ru spins order magnetically above say 90 K with a ferromagnetic component within at 5 K. As far the value of higher field (> 4 T) saturation moment is concerned, one cannot without ambiguity extract the value for Ru contribution, because the contribution from Cu cannot be ignored. In an under-doped HTSC compound Cu contributes an unknown paramagnetic signal to the system.

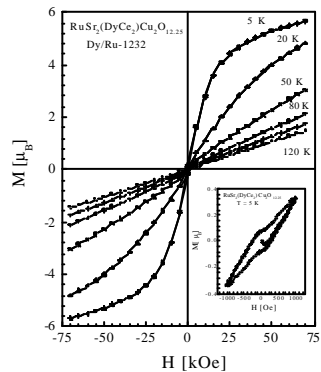

**Figs. 33** (*M*) vs. applied field (*H*) plots for Dy/Ru1232 sample in higher applied fields.

Figure.34 shows the isothermal magnetization (*M*) vs. applied field (H) behaviour for Dy/Ru-1232 in low fields of 1000 Oe = H = 1000. Clear M vs. H loops are seen at 5, 20, and 50 K, but not at 80 K. The applied fields are in the range of -1000 Oe = H = 1000. The returning moment ($M_{rem}$) i.e. the value of magnetization at zero returning field and the coercive filed ($H_c$) i.e. the value of applied returning field to get zero magnetization are clearly seen up to 50 K. Worth mentioning is the fact that Dy (magnetic rare earth) in the compound must order magnetically below 0.5 K and Ce is known to be in tetravalent non-magnetic state hence the $M_{rem}$ and $H_c$ arising from the ferromagnetic hysteresis loops do belong to Ru only. Hysteresis loops are not seen for M vs. H plots at or above 80 K. For various hysteresis loops being observed from M vs. H plots below 80 K, the values of both $M_{rem}$ and $H_c$ decrease with increase in T. Both $M_{rem}$ and $H_c$ being observed for Ru-1232 are much higher than reported for other magneto-superconductor

Ru-1212 and comparative to Ru-1222. For Ru-1212 the hysteresis loops are reported quite narrow. This indicates that in Ru-1232 the ferromagnetic domains are less anisotropic and more rigid like Ru-1222 and unlike to that for Ru-1212. Worth mentioning is the fact that the Ln/Ru-1232 compounds are not yet synthesized in pure phase, see Figure 5. Hence their magneto-superconductivity can yet not be conclusive. In any case the preliminary unpublished results on Ln/Ru-1232 compounds are shown above in the present review of rutheno-cuprates.

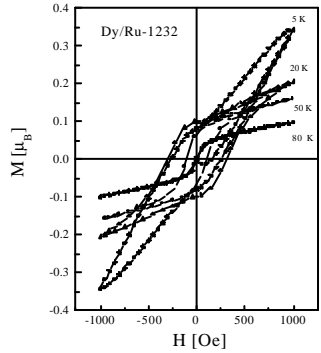

**Figs. 34** (*M*) vs. applied field (*H*) plots for Dy/Ru1232 sample in low applied fields

**3.6 The complex magnetic ordering of Ru in Ru-1212 and Ru-1222**

It is discussed in the beginning of the review that the exact nature of the ordering of Ru moments in various rutheno-cuprate magneto-superconductors is not yet well understood. In fact recent magnetization results do conflict with each other viz. ref. [10,15,16]. Also the magnetization picture [15,16] is not the same as proposed via neutron scattering studies [11,12,31]. Most recent results on both Ru-1222 [27,45] and Ru-1212 [46] have provided even more complex picture of the magnetism of these compounds. Various magnetization measurements on Ru-1222 have provided the evidence for magnetic phase separation [27]. Also it was pointed out that magnetic structure of $RuO_6$ octahedra in Ru-1222 is different than that for Ru-1212 [27]. Here some of our very recent results [45,46] pertaining to the AC susceptibility measurements on both Ru-1222 and Ru-1212 are given. A very powerful technique to evidence spin-glass (SG) behavior is the ac susceptibility $\chi_{ac}$ measurements. It is expected to a spin-glass system that both components $\chi'$ and $\chi''$ of $\chi_{ac}$ present a sharp, frequency dependent cusp. The position of the cusp $\chi'$ defines the freezing temperature $T_f$, which coincides with the inflection point in $\chi''$. It is also well known that dc magnetic fields as low as a few hundreds of Oersted can round this cusp up. In Fig. 35 we present the ac susceptibility for our sample measured at H = 50 Oe. The main panel of Fig. 34 presents the temperature dependence of $\chi'$ for 4 different frequencies (10,100, 1000, and 10000 Hz). $\chi'$ presents a sharp, frequency dependent peak at $T_f \sim 72\ K$ for all Ln/Ru-1222 samples. The peak shifts to lower temperatures and its intensity increases as the frequency of the excitation field is decreased. Interestingly for Y/Ru-1212 sample the frequency dependence of $\chi'$ is not seen.

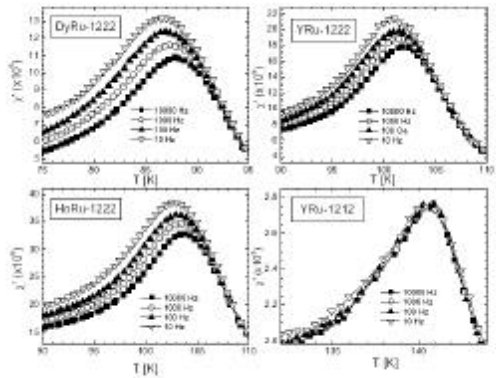

**Fig. 35.** Real part of the ac susceptibility as a function of temperature for H = 50 Oe and four different frequencies, for all four samples. The peak position defines the freezing temperature $T_f$ (for the Ln/Ru – 1222 samples).

These results indicate that though the Ru-1222 compounds possess SG behaviour of Ru moments [45], the Ru-1212 system does not [46]. The presence/absence of SG component in these systems was shown by various other measurements viz. TRM (thermoremanent magnetization) and IRM (isothermal remanent magnetization) [45,46]. A moot question arises, why the two systems with similar $RuO_6$ octahedra do have different magnetic structure of Ru moments?. The answer perhaps lies with the fat that in Ru-1222 relatively larger variation of oxygen content is permitted than in Ru-1212. This implies necessary that $RuO_6$ becomes $RuO_{6-d}$ with varying $Ru^{4+}/Ru^{5+}$ valence states.

### 3.6 Electrical transport properties: PPMS results

**A. Ru-1212 samples**

Figure 36 shows the resistance ($R$) *versus* $T$ for an as-synthesized (superconducting) $RuSr_2GdCu_2O_{8-\delta}$ sample in zero, 3 and 7 T applied fields. The $R$-$T$ behaviour without any applied magnetic field is metallic down to 150 K and semiconducting between 150 K and 25 K, with a superconducting transition onset ($T_c^{onset}$) at 25 K and $R = 0$ at 20 K. This behaviour is typical of underdoped HTSC compounds. Also observed is an upward hump ($T_{hump}$) in $R$-$T$ around 140 K, which indicates the possibility of antiferromagnetic ordering of spins. The $R$-$T$ behaviour under an applied field of 7 T is nearly the same above $T_c^{onset}$, except that $T_{hump}$ is completely smeared out due to possible change in the magnetic structure. Also in 7 T applied field the $T_c^{onset}$ decreases to around 10 K and $R = 0$ is not observed down to 5 K.

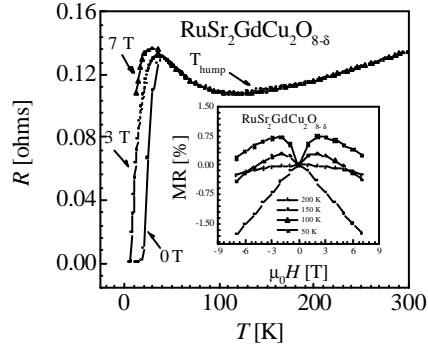

**Fig. 36**. R-T plot and magneto-transport behaviour for an as-synthesized (superconducting) $RuSr_2GdCu_2O_{8-\delta}$ sample.

In an intermediate field of 3 T, both $T_c^{onset}$ and $T_c^{R=0}$ decreased, to 20 and 10 K respectively. For conventional HTSC, $T_c^{onset}$ remains nearly the same under all possible applied fields, with decreasing $R = 0$ temperature and an increased transition width ($T_c^{onset} - T_c^{R=0}$). Therefore, a different type of broadening of the transition under a magnetic field is obtained for Ru-1212 from that reported for conventional HTSC. In earlier reports on Ru-1212, the $T_c^{onset}$ under a magnetic field decreased like the present case. The present behaviour of transition broadening under a magnetic field is presumably due to formation of SNS/SIS junctions/clusters in the present and similar samples. Non-superconducting $RuSr_2GdCu_2O_{8-\delta}$ might be stacked between superconducting $Ru_{1-x}Cu_xSr_2GdCu_2O_{8-\delta}$, resulting in ideal SIS or SNS junctions within the material.

Inset of Fig. 36 shows the magnetoresistance (MR) behaviour of the present Ru-1212 sample at various fields and temperatures. MR is negative in all applied fields upto 7 T above the magnetic ordering temperature, *i.e.* at 150 K and 200 K. Maximum negative MR of up to 2 % is observed at 150 K, which is close to the magnetic ordering temperature of around 140 K. At temperatures below the ordering temperature (100 K and 50 K), MR displays a positive peak at low fields and becomes negative at higher fields. This behaviour is in general agreement with previous reports.

The R-T plot and magneto-transport behaviour for a 100-atm $O_2$-annealed Ru-1212 sample revealed interestingly no $R = 0$ and only a $T_c^{onset}$ was observed around 20 K. Other characteristics in terms of $T_{hump}$ in R-T at around 140 K and the systematic changes in MR with applied field and T were the same as for the as-synthesized sample.

**B. Ru-1222 samples**

Figure 37 depicts the $R$-$T$ plots in 0 and 7 T fields for an 100-atm $O_2$-annealed Ru-1212 sample. The $R$-$T$ behaviour in zero field is similar to that observed for the as-synthesized sample with some improvement towards metallic conductivity. Superconductivity starts with $T_c^{onset}$ at 51 K and the $T_c^{R=0}$ is seen at 43 K.

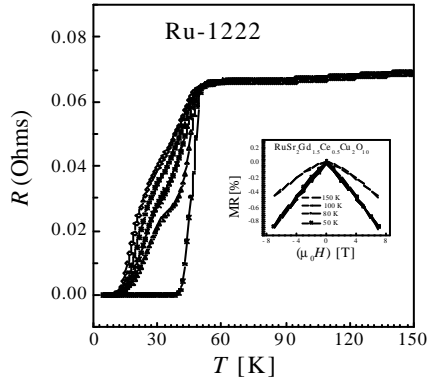

**Fig. 37.** *R-T* plot and magneto-transport behaviour for a 100-atm $O_2$-annealed $RuSr_2(Gd_{0.75}Ce_{0.25})_2Cu_2O_{10-\delta}$ sample. The inset shows the MR behaviour of the same.

The *R-T* behaviour under an applied field of 7 T is nearly the same as that of 0 T above $T_c^{onset}$. However in 7 T field $T_c^{R=0}$ is observed only around 12 K. In intermediate fields of 1 T, 3 T and 5 T, $T_c^{onset}$ remains nearly invariant and $T_c^{R=0}$ are seen at 18 K 16 K, and 14 K, respectively. Also seen is a shoulder in the *R-T* curve in all applied fields, the origin of which is not known. When these results of magneto-resistivity are compared with results for the as-synthesized sample, one finds that $T_c^{R=0}$ is nearly doubled from 23 K to 43 K after the 100-atm $O_2$ annealing. Furthermore, $T_c^{R=0}$ is observed in 7 T field also, which is not the case for the as-synthesized sample. Magneto-resistivity results for the 100-atm $O_2$-annealed Ru-1212 sample substantiate the magnetization results, indicating that superconductivity is enhanced upon the 100-atm $O_2$ annealing.

In the inset of Fig. 37, the MR data of the same Ru-1222 sample is shown at various temperatures and fields, revealing a small negative MR effect in the whole temperature range. Below 100 K the degree of MR is nearly the same in all applied fields and the nature of the MR effect is of the tunnelling-magneto-resistance (TMR) type as judged from the curve shape. Also note that the MR behaviour of the present Ru-1222 sample is different from that of Ru-1212, section 3.6 (A). Ru-1212 exhibited systematic changes in sign of MR at various *T* and fields.

**C. Magneto-transport of $N_2$-annealed non-superconducting Ru-1222 sample**

Figure 38 depicts the resistance versus temperature (R-T) behavior for $N_2$-annealed Ru-1222 sample in magnetic fields of 0, 3 and 6 Tesla. The R-T behavior of this compound is semiconducting down to 2 K. No superconducting transition is observed in the whole temperature range studied (2-300 K). Further, in low temperature region, an appreciable MR is seen for the $N_2$-annealed sample. Magneto-resistance (MR), as a function of applied field, at temperatures of 5 and 10 K, for $N_2$-annealed sample, is plotted in inset of Fig.3. MR of >20 % is observed at 5 K in an applied field of up to 9 Tesla. At 2 K, around 20% MR is seen even in low applied field of 3 Tesla (plot not shown).

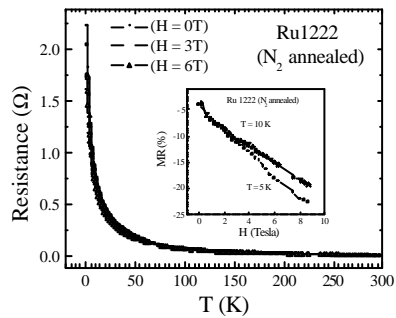

**Fig. 38.** Resistance (R) vs. temperature (T) plots in 0, 3, and 6 T applied magnetic fields for $N_2$-annealed Ru-1222. Inset shows the Magnetoresistance (MR%) at 5 K in applied fields up to 9 Tesla.

## 4. SUMMARY AND CONCLUDING REMARKS

We reviewed our various results on rutheno-cuprate magneto-superconductors $RuSr_2GdCu_2O_{8-\delta}$ (Ru-1212) and $RuSr_2(Gd_{0.75}Ce_{0.25})_2Cu_2O_{10-\delta}$ (Ru-1222) synthesized by NPHT process. Also reviewed are the results for HPHT synthesized various Ln/Ru-1212, 1222 and 1232. Various important results being reported by other international groups are incorporated and discussed critically. It is observed, that it is difficult to control the oxygen content of Ru-1212, though the same is possible up to some extent for Ru-1222. Samples of both phases exhibited superconductivity, presumably in the $CuO_2$ plane at low temperatures coexisting with magnetic ordering of Ru spins of predominantly antiferromagnetic type below 140 K having a ferromagnetic component appearing below 20 K. Electrical conductivity measurements indicate that the $RuO_{2-\delta}$ layer takes part in conduction besides the $CuO_2$ plane. The magnetic ordering temperature of Ru spins is seen as a clear hump in the resistivity measurements, establishing the magnetic spins interaction with the conduction carriers.

The $Ru_{0.9}Sr_2YCu_{2.1}O_{7.9}$ being synthesized by HPHT (high pressure high temperature) solid-state reaction route exhibits superconductivity below 30 K. Also the Ru-spins are ordered magnetically above 143 K, with a ferromagnetic component at 5K. $Ru_{0.9}Sr_2YCu_{2.1}O_{7.9}$ shows clear diamagnetic transitions in zero-field-cooled (ZFC) magnetic susceptibility in applied fields of up to few hundred Oe (< 300 Oe) at superconducting transition temperature ($T_c$), followed by a dip in the field-cooled (FC) magnetization at same temperatures with near saturation of the FC signal below 10 K. Though the dip in FC magnetization at $T_c$ followed by saturation below 10 K indicates towards good volume fraction of superconductivity, the ZFC transition is not saturated down to 5 K. This suggests that though superconductivity is of bulk nature, the same is not well connected to provide sufficient surface shielding currents for ZFC process. Interestingly the compound does not exhibit R = 0 state down to 5 K. Low field (<1000 Oe) *M* vs. *H* plots show clearly that both superconducting and the ferromagnetic components are present in the compound at 5 K. The sample shows ferromagnetic like hysteresis loops at 5, 20 K in *M* vs. *H* plots. Though the complete saturation of moments is not achieved at applied fields of as high as 7 T, the near saturation values observed at 7 T and 5 K is 1.17µB. This value is higher than as expected theoretically for magnetic ordering of low spin (1/2) states of $Ru^{5+}$ and considerable less than for high spin (3/2) state. The returning moment value as seen from hysteresis loop is 0.079 µB per Ru.

The Ln/Ru-1222 materials become magnetically ordered at $T_M$ =152(2) K regardless of Ln [32,47]. The wide ferromagnetic-like hysteresis loops which open at 5 K, close themselves around $T_{irr}$= 90-100 K and the remanent magnetizations ($M_{rem}$) and the coercive fields ($H_C$) become zero. Surprisingly, at $T_{irr}$<T< $T_M$ a reappearance of the $M_{rem}$ and $H_C$ (with a peak at 120-130 K) is observed for all three samples studied. For the non-magnetic Ln=Y compound, the extracted saturation moment at 5 K and the effective paramagnetic moment are is 0.75 and 2.05 $\mu_B$/Ru, values which are close to the expected 1 $\mu_B$ and 1.73 $\mu_B$ respectively, for the low-spin state of $Ru^{5+}$. We argue that the Ru-1222 system becomes (i) anti-ferromagnetically (AFM) ordered at $T_M$. In this range a metamagnetic transition is induced by the applied field (ii). At $T_{irr}$ < $T_M$, weak-ferromagnetism (W-FM) is induced by the canting of the Ru moments.

Two most important issues related to rutheno-cuprates are about the phase purity of these compounds [48], and the discussion of ensuing basic physics [49] related to them. This is some thing, which need to be debated for any new material being invented. It is reported that $RuSr_2GdCu_2O_8$ decomposes under high-temperature treatment [48], giving rise to micro-islands of the melt depleted Ru and Cu-enriched phase. Interestingly enough Ru depleted phase viz. $Ru_{1-x}Sr_2GdCu_2O_8$ could invoke for superconducting but not magnetic material. For example composition like $Ru_{1-x}Sr_2GdCu_{2+x}O_8$ could show superconductivity with say x = 0.5 but not magnetic characteristics [50]. These doubts regarding phase separation in rutheno-cuprates were casted earlier also [25,29,33]. Recent reports [48,50] once again have asked for the phase purity of widely discussed Gd/Ru-1212 compounds. As discussed in introduction itself, the phase purity issue of rutheno-cuprates is still far from conclusive. It seems the magnet-superconductivity of rutheno-cuprates is yet in pre-mature stage and a lot more need to be done before concluding the physical properties of these materials.

## 5. ACKNOWLEDGEMENT

The vast amount of work being given in the current review would not have been possible beside the help of several collaborators during my visits to various international laboratories. The work on rutheno-cuprates was first started during my visit to Max-Planck Institute, Stuttgart Germany, for which the kind help and guidance from Prof. E. Gmelin and Dr. R.W. Henn is acknowledged. Later work was continued in Tokyo Institute of Technology (TITECH), Japan under the able guidance of Prof. H. Yamauchi and Prof. M. Karppinen. Several HPHT phases of rutheno-cuprates were synthesized and characterized in National Institute of Material Science (NIMS) Japan under guidance and encouragement by Dr. E. Takayama-Muromachi. Finally coming back to India now Prof. S.K. Malik from Tata Institute of Fundamental Research (TIFR) Mumbai is guiding and helping in further developing the field of rutheno-cuprate magneto-superconductors.

There are also some collaborators, being time to time involved for specific studies on rutheno-cuprates. They include Prof. R.S. Liu from Hsinchu, Taiwan, Prof. Oscar de Lima and Dr. Claudio A. Cardossso from Instituto de

Fisica, UNICAMP, Brazil and Prof. I. Felner from Racah Institute of Physics Jerusalem, Israel. Infact Prof. I. Felner has been continuously a guiding spirit behind the work on rutheno-cuprates being presented in the current review.